\documentclass[11pt,letterpaper]{article}
\usepackage[utf8x]{inputenc}
\usepackage{jcappub}
\usepackage{blkarray, bigstrut}
\usepackage{epsfig,latexsym,cancel,amssymb,amsmath}
\usepackage{hyperref}
\usepackage{graphicx}
\usepackage{caption}
\usepackage{subcaption}
\usepackage{comment}
\usepackage{xcolor}
\usepackage{hyphenat}
\usepackage{bm}
\usepackage[normalem]{ulem}
\usepackage{float}

\def\beq{\begin{equation}}
\def\eeq{\end{equation}}
\def\bea{\begin{eqnarray}}
\def\eea{\end{eqnarray}}

\emergencystretch=\maxdimen
\def\Gaia{\emph{Gaia}}
\begin{document}
\title{Astrometric Microlensing of Primordial Black Holes with \Gaia{}}
\author[a]{Himanshu Verma,}
\author[a]{Vikram Rentala}

\affiliation[a]{Department of Physics, Indian Institute of Technology Bombay, Powai, Mumbai, Maharashtra, 400076, India}
\emailAdd{verma.himanshu002@gmail.com}
\emailAdd{rentala@phy.iitb.ac.in}

\abstract{The \Gaia{} space telescope allows for unprecedented accuracy for astrometric measurements of stars in the Galaxy. In this work, we explore the sensitivity of \Gaia{} to detect primordial black hole (PBH) dark matter through the distortions that PBHs would create in the apparent trajectories of background stars, an effect known as  astrometric microlensing~(AML). We present a novel calculation of the lensing probability, and we combine this with the existing publicly released \Gaia{} eDR3 stellar catalog to predict the expected rate of AML events that \Gaia{} will see. We also compute the expected distribution of a few event observables, which will be useful for reducing backgrounds.
Assuming that the astrophysical background rate of AML like events due to other sources is negligible, we then compute the potential exclusion that could be set on the parameter space of PBHs with a monochromatic mass function. We find that \Gaia{} is sensitive to PBHs in the range of $0.4~M_\odot$ - $5\times10^7~M_\odot$, and has peak sensitivity to PBHs of $\sim 10~M_\odot$ for which it can rule out as little as a fraction $3\times10^{-4}$ of dark matter composed of PBHs. With this exquisite sensitivity, \Gaia{} has the potential to rule out a PBH origin for the gravitational wave signals seen at LIGO/Virgo. Our novel calculation of the lensing probability includes for the first time, the effect of intermediate duration lensing events, where the lensing event lasts for a few years, but for a period which is still shorter than the \Gaia{} mission lifetime. The lower end of our predicted mass exclusion is especially sensitive to this class of lensing events. As and when time-series data for \Gaia{} is released, and once we have a better understanding of the astrophysical background rate to AML signals, our prediction of the lensing rate and event observable distributions will be useful to estimate the true exclusion/discovery of the PBH parameter space utilizing this data.
 }

%\pacs{ }
\maketitle

\section{Introduction}

Dark matter (DM) makes up 25\% of the energy density of the universe today, yet it is still unknown as to what constitutes the DM. One promising theory is that DM is made up of Primordial Black Holes (PBHs)~\cite{1975Natur.253..251C, Khlopov:2008qy, Villanueva-Domingo:2021spv}. Unlike astrophysical black holes which form due to the collapse of stars at the end of their life cycle, PBHs are expected to form in the early universe from excess density perturbations that gravitationally collapse while overcoming resistance from pressure~\cite{1971MNRAS.152...75H}. In particular, this could happen due to either an excess in the primordial power spectrum on small scales~\cite{Garcia-Bellido:1996mdl,Clesse:2015wea, Choi:2022btl}, or due to sudden drops in pressure in the matter-radiation fluid~\cite{Garcia-Bellido:2017fdg}. For other formation channels for PBHs see for e.g.~\cite{Carr:2020xqk}.

There are a number of observational constraints on PBHs that make up a significant fraction of the DM (see ref.~\cite{Carr:2020gox, Green:2020jor} for overviews of these constraints). Usually, these constraints are phrased assuming that the PBHs have a monochromatic mass function -- where a fraction $f$ of the DM density is assumed to be made up of PBHs of mass $M$. The most stringent method of setting constraints on PBHs varies depending on the assumed value of $M$. For $M$ of $10^{-11} ~M_\odot$ and above, a combination of constraints from microlensing~\cite{Oguri:2017ock,Croon:2020ouk,Green:2020jor,MACHO:2000nvd,Zumalacarregui:2017qqd,Niikura:2019kqi}, gravitational wave signatures~\cite{Kavanagh:2018ggo, LIGOScientific:2019kan, Chen:2019irf}, accretion~\cite{Brandt:2016aco, 2014ApJ...790..159M} and dynamical effects~\cite{Serpico:2020ehh,Hektor:2018qqw,Manshanden:2018tze,Lu:2020bmd}, seem to rule out a fraction $f\gtrsim 0.1$ of the DM as being made up of PBHs. However, it is possible that PBHs in this mass range make up a smaller, sub-dominant component of the DM.

A large number of experiments have leveraged the technique of \textit{photometric microlensing} (PML) to set observational constraints on the PBH abundance~\cite{Niikura2019, Smyth:2019whb, Niikura:2019kqi, Griest:2013aaa, Griest:2013esa, EROS-2:2006ryy, MACHO:2000nvd}. PML relies on the prediction of general relativity that light bends around massive objects. If PBHs constitute the DM then they are distributed throughout galaxies. If a star passes behind a PBH, then the PBH acts like a gravitational lens, and the light from such a star would get bent and focused, leading to a temporary brightening of such a star. Thus, surveys which have precision photometric sensitivity, i.e. sensitivity to small brightness variations, have been employed very fruitfully to set very strong constraints on the existence of PBHs in the mass range $10^{-11}-30 ~M_\odot$.

In addition to the apparent magnification of a star due to a PBH in its foreground, another consequence of lensing is that the apparent trajectory of the star would appear to be distorted relative to its true trajectory. Such an effect is called \textit{astrometric microlensing} (AML)~\cite{1995A&A...294..287H,1995ApJ...453...37W,1995AJ....110.1427M,Gould:1996nb}. Surveys which have precision astrometric sensitivity, i.e. sensitivity to the position and velocity of the star, would be able to detect such an effect. In principle AML should be more sensitive than PML to the presence of PBHs. This is because in AML the relevant observable, which is the deflection of a star's apparent position from its true position, falls off as $1/\theta$, where $\theta$ is the angle between the true position of the lens and the star on the celestial sphere~\cite{2000ApJ...534..213D}. This is in contrast to PML where the relevant observable is the change in magnification, and this falls off faster, as $1/\theta^4$~\cite{2000ApJ...534..213D}.

In the past few years, AML techniques have already started to be employed to detect PBH and PBH-like candidates. Recently, there has been an exciting claim of discovery of an isolated stellar-mass black hole lens using the AML technique with HST data~\cite{OGLE:2022gdj}. AML searches have also been employed to detect signatures of intermediate mass black holes in the mass range $10^2-10^5~M_\odot$~\cite{2016MNRAS.460.2025K, Kains:2018vnd}. AML observations have also been proposed as follow ups for PBH candidates detected via PML methods in order to measure the properties of the lens with greater precision~\cite{Wyrzykowski:2015ppa, 2016ApJ...830...41L, Han_2019}. Besides PBHs, AML can be used to set constraints on exotic ultra-compact mini-clusters halos (UCMCHs)~\cite{Li:2012qha, Malbet:2021rgr}, and also on more conventional DM sub-halos~\cite{Erickcek:2010fc, VanTilburg:2018ykj}.

What is needed for precision astrometry is highly accurate position measurements and also high cadence. The more accurate the position measurements, the more easily small deflections in the trajectory of a star can be detected. High cadence ensures that the star's position is tracked frequently enough that a lensing event is not missed. Both these factors affect sensitivity to lower mass PBHs. At the other extreme, for high mass PBHs, the lensing event durations can be extremely large, and if the mission life time is small compared to a typical event duration, then the detection of a distortion in the trajectory of a star can be difficult.

The \Gaia{} satellite, launched by ESA, has been surveying the sky since 2014. \Gaia{} is an optical all-sky survey satellite with a large field-of-view $\sim0.72^\circ \times 0.69^\circ$~\cite{2012A&A...538A..78L}. The data collected by this satellite provides an exciting opportunity for high precision astrometry. \Gaia{} is expected to observe over 1 billion stars and obtain sub milli-arcsecond level precision position measurements in a single pass. Already, it has provided us with unprecedented views of the Milky Way. However, at the moment, time-series data, which would enable us to make high precision searches for AML events, has not been publicly released by the collaboration.

In this work, we set out to estimate the sensitivity of \Gaia{} to PBHs through their expected AML signatures. We first make a prediction for the expected rate of AML events that \Gaia{} will detect for a given assumption of PBH parameters $f$, $M$. We argue that lensing  events can be classified into one of three types: short, intermediate, and long duration lensing events (SDLEs, IDLEs and LDLEs). Our prediction for the lensing rate for these classes of events involves a novel calculation of the lensing probability of a given background star. We combine this probability calculation with a catalog of stars for which \Gaia{} is expected to have high precision astrometric data, to obtain a lensing event rate prediction. Besides the total rate, we also make predictions for the distribution of expected event observables, such as location on the sky, event duration, maximum deflection angle, etc. These predictions will be useful for characterizing the space of observables for genuine signal events which can distinguish them from backgrounds.

There are three main types of backgrounds that \Gaia{} will see. The first type are statistical fluctuations in the centroiding of the camera image of the background star. The second type are instrumental systematics. The third type are astrophysical backgrounds which can mimic PBH induced AML signatures. For example, besides AML events due to PBHs, \Gaia{} is also expected to see AML events due to microlensing caused by foreground stars~\cite{2002MNRAS.331..649B, 2018A&A...618A..44B}, objects formed from stellar collapse~\cite{2020ApJ...889...31L, 2021MNRAS.508.3877G, 2022A&A...666L..16J}, brown dwarfs~\cite{Belokurov:2001vh, 2018A&A...620A.175K}, and free floating planets~\cite{2018Ap&SS.363..153H}. Another type of astrophysical background could be due to stars which do not follow rectilinear motion. Isolated stars are expected to approximately follow rectilinear motion in the Galactic rest frame. However, unresolved binary companions or local gravitational potential gradients, may introduce distortions in the observed trajectories compared to the trajectories expected from rectilinear motion. These distortions could also be incorrectly interpreted as AML events, and could therefore lead to an additional source of background to PBH searches.

If we could characterize the space of observables of these backgrounds, then we could compare it to the predicted space of observables for genuine lensing signals, and use it to reduce backgrounds --  for example by cutting out regions of parameter space that are unlikely to occur due to PBH induced AML signals. The detailed backgrounds are hard to quantify precisely without full numerical simulations to obtain both their rate and observable distribution. However, we attempt to give some analytic estimates of the statistical background rate for IDLEs. We also try to identify the astrophysical backgrounds that are likely to be important for PBH searches. The degree to which such astrophysical backgrounds can be further reduced by examining the distribution of event observables is an important question that needs further investigation, but is beyond the scope of the present work.

In order to predict the regions of PBH parameter space that are likely to be discovered/excluded with time-series data from \Gaia{}, we need a computation of both the signal rate (which we have done), and of the background rate (which needs further study). We make the simplified assumption of a negligible background rate to calculate the expected exclusion on the PBH parameter space. While the time-series data for \Gaia{} has already been collected, it has not yet been publicly released. We expect that our prediction of the lensing signal rate and the signal event observable distributions will prove useful to analyze this data, as and when it becomes available, in order to compute the true exclusion bounds.

With our assumption of a negligible background rate, we find that \Gaia{} is most sensitive between $0.4~M_\odot$ - $5\times10^7~M_\odot$, with peak sensitivity to masses $\sim 10~M_\odot$, where fractions $f$ as low as $3\times10^{-4}$ of the DM relic density can be constrained. Our work is the first attempt to make such a prediction for \Gaia{} by looking at deflections of the trajectories of individual stars. Our work is enabled in part by the existing \Gaia{} eDR3 data release~\cite{2021A&A...649A...1G}, which already gives an indication of the number of stars that \Gaia{} is able to track, and also lists their measured properties that are relevant for calculating the AML rate.

Our work builds upon the work of Dominik and Sahu~\cite{2000ApJ...534..213D} who estimated lensing probabilities for \Gaia{}. However, in this work the authors did not attempt to calculate the number of lensing events that \Gaia{} would see or set an exclusion limit on the PBH parameter space. Moreover, in their work, the authors only considered LDLE type events, with event durations longer than the \Gaia{} mission lifetime. In our work, we show for the first time the importance of IDLE type events, which have event durations of a few months to a few years, a time-scale which is intermediate between \Gaia{}'s sampling time and its mission lifetime. As we will show, searches for IDLE type events are important for sensitivity to relatively lower mass PBHs with masses as low as $\sim 0.1~M_\odot$\footnote{SDLE type events have event durations shorter than \Gaia{}'s sampling time and will therefore show up as a fluctuation in at most a single pass observation of a background star. However, as we shall argue in sec.~\ref{sec:background_stat}, SDLEs will be completely dominated by statistical background due to centroiding uncertainty. This will therefore render SDLEs unusable in practice for detecting PBHs.}.

We now provide some broader context for the impact of the work that we present in this paper.  For one, an exclusion of PBHs with $\mathcal{O}(10)~M_\odot$ mass as making up less than a fraction $f\sim 10^{-3}$ of the DM would rule out the possibility that the mergers of binary black holes being seen at LIGO/Virgo~\cite{LIGOScientific:2018glc} are due to PBHs~\cite{Carr:2020xqk}. If \Gaia{} were to \textit{discover} even a fraction of the DM in the form of PBHs in the mass range of  $10^{-1}-10^7 ~M_\odot$, that we have claimed the survey is sensitive to, then this would also have profound implications. If PBHs in this mass range constitute even a small fraction of the DM density, it would rule out weakly interacting massive particles (WIMPs) as a candidate for the rest of the dark matter. Refs.~\cite{Lacki:2010zf, Boucenna:2017ghj, Adamek:2019gns, Kadota:2021jhg} have argued that annihilation of WIMP DM would be greatly enhanced due to accretion around PBHs in this mass range, and this would lead to a large indirect detection signal which is incompatible with present limits, for a PBH density fraction $f>10^{-9}$.
A technical implication of our work is that our novel lensing probability calculation can be easily adapted to predict the rate of lensing events due to UCMCHs/DM sub-halos, or ordinary star-star lensing.

This paper is organized as follows. In sec.~\ref{sec:Gaia}, we describe the properties of the \Gaia{} satellite that are relevant for AML and we also discuss the \Gaia{}~eDR3 data release from which we form a model of the stellar distribution in the Galaxy. In sec.~\ref{sec:LensingProb}, we describe the expected signals due to PBH induced AML events and we provide analytic expressions for the event durations and lensing probabilities for stars in the Galaxy. We also introduce the distinction between SDLE, IDLE, and LDLE types events in this section. In sec.~\ref{sec:lensingrate}, we combine the lensing probability calculation with the Galactic stellar model to numerically compute the expected total number of lensing events that \Gaia{} will see over its mission lifetime, and we also compute the expected distribution of some simplified AML observables. Next, in sec.~\ref{sec:background}, we describe various backgrounds to PBH searches. We also describe which backgrounds we expect to be important, and some background reduction techniques. In sec.~\ref{sec:Results}, assuming negligible backgrounds, we then calculate an expected exclusion curve on the PBH parameter space. In sec.~\ref{sec:discussion}, we discuss some key assumptions of our rate estimation and how the robustness of our exclusion curve can be improved by future numerical studies.
Finally, we summarize and conclude in sec.~\ref{sec:Summary}. In the appendix, we derive some analytic expressions for the event observables.

\section{The \Gaia{} telescope and stellar catalog}
\label{sec:Gaia}
\emph{Gaia} is a space-based optical-telescope designed to obtain precise positions and velocities of stars in our Galaxy. \emph{Gaia} started surveying the sky in July, 2014 and has already surpassed its expected 5 year mission lifetime. It is expected that \Gaia{} will provide time-series data for astrometric position, proper motion, and parallax for more than a billion stars~\cite{2016A&A...595A...1G}. With such exquisite time-series data, \Gaia{} has the potential to be sensitive to distortions in the trajectories of stars due to astrometric microlensing events. \Gaia{} has not yet publicly released time-series data, although this data is expected to be released in the future~\cite{2021A&A...649A...1G}. Thus, at the moment we can not directly analyze \Gaia{} data for potential AML signals.

The goal of our work is two fold. First, we would like to estimate the expected sensitivity of \Gaia{} to AML signals due to PBH induced lensing of background stars. Second, we would like to predict the distribution of event observables such as location on the sky, maximum astrometric deflection angle, and event durations, for different assumptions of the PBH mass and abundance. As and when time-series data of \Gaia{} is made available, we expect that our work will prove useful when analyzing this data.

In this work, we utilize information from \Gaia{} in two distinct ways. First, we make use of the known detector properties, such as how often \Gaia{} scans a particular region of the sky and the uncertainty on the astrometric position of a star in a single pass, to estimate the probability of \Gaia{} detecting a lensing signature. Second, although the \Gaia{} collaboration has not released \textit{time-series} data, they have already made several public releases of high precision, \textit{time-averaged} astrometric positions, proper motion, and parallaxes of nearly 1.5~billion stars~\cite{2021A&A...649A...1G}.
This catalog already gives us the most precise map of the Milky~Way. We will therefore use this catalog to predict the event rate for AML signals.

In sec.~\ref{subsec:GaiaDetProperties}, we discuss the detector properties that will be relevant for estimating the lensing probabilities. In sec.~\ref{subsec:GaiaEDR3Catalog}, we will then discuss the \Gaia{} data release and how we build a stellar catalog from the data to predict the expected AML event rate and event observable distributions.

\subsection{\Gaia{} detector properties}
\label{subsec:GaiaDetProperties}
\Gaia{} is located at the Lagrange point L$_2$. The satellite has two telescopes with independent fields-of-view, each of angular size $0.72^\circ \times0.69^\circ$~\cite{2012A&A...538A..78L}, with an angular separation of $106.5^\circ$ between the two fields-of-view. \Gaia{} spins about an axis inclined at an average angle of 45$^\circ$ with respect to the line joining the detector and the sun with an angular speed of $60 \textrm{ arcsec/sec}$. This rotation axis precesses at a rate of $5$ revolutions per year. \Gaia{} also executes a complicated motion around the Lagrange point L$_2$. As the satellite spins about its rotation axis, the fields-of-view sweep out different regions of the sky. The precession of the rotation axis ensures that after a single rotation the satellite is scanning a slightly different patch of the sky. Given the precession rate and the spinning rate, it is expected that on average a given star would be seen once in each field-of-view as \Gaia{} rotates, and would then drop out of the fields-of-view. The observation of a star in a single telescope during a single rotation constitutes one ``pass''. The advantage of the two fields-of-view is that this allows \Gaia{} to perform \textit{global} astrometric measurements of the position of a star relative to other background stars in another field-of-view. Thus, in a single pass, \Gaia{} will be able to make one observation of the astrometric position of a star which can later be assigned a specific global astrometric co-ordinate (say, \{$\alpha, \delta$\} in Galactic co-ordinates).

\emph{Gaia}'s motion results in a complicated scanning law~\cite{2016A&A...595A...1G} which determines the time-interval between successive observations of the same star. The time-interval between successive observations is also not constant. Over an observational time scale of $t_\textrm{obs} = 5$~years, \Gaia{}
is expected to observe each star approximately 70 times (35 times in each field-of-view). Thus, at the end of the observational period \Gaia{} is expected to have non-uniformly sampled time-series data with about 70 astrometric positions for each star. For simplicity, we will treat the two successive observations of a star in each field-of-view as effectively a single pass. We then take the time-interval between these effective passes to be uniform, with a spacing $t_s = 5 \textrm{ years}/35 \approx 52.2 \textrm{ days}$\footnote{In principle, the observation of a star in the two different fields-of-view can be considered to be two separate passes and could potentially be used to search for short duration microlensing events with time scales of a few hours. However, it seems unlikely given \Gaia{}'s sampling, that such events would occur at a significant enough rate to be detectable for realistic PBH distributions, so we will not discuss this possibility further in this work.}.

\Gaia{} is equipped with three color bands, white light G-band (330-1050 nm), Blue prism Photometer (BP-band: 330-680 nm), and Red prism Photometer (RP-band: 640-1050 nm). It is sensitive to stars as faint as 21 G-band magnitude. The uncertainty on the astrometric position of a star in a single pass depends upon the apparent brightness. In order to model the uncertainty, we could use use a fitted model for this as a function of $m_G$ (G-band magnitude) given in ref.~\cite{2018A&A...618A..44B}. This model is based on the Monte Carlo centroiding simulations done in ref.~\cite{2018MNRAS.476.2013R}. Rather than a direct measurement of the uncertainty on the absolute position of a star in a single pass, the model reports different uncertainties across (AC) and along (AL) the scanning direction of \Gaia{}, as a star passes through the spinning fields-of-view of each telescope. These fitted uncertainties are given by the formula,
\begin{align}
\label{eq:AsigmaSim}
    \widetilde{\sigma}_a = & \frac{1}{\sqrt{2}}\begin{cases}
     \begin{cases}
        0.200+0.483 e^{0.690(m_G-12)} \textrm{ mas} & \textrm{\, for \,} m_G \leq 13, \\
        1.140+0.420 e^{0.771(m_G-12)} \textrm{ mas} & \textrm{\, for \,} m_G > 13
    \end{cases} & ; \textrm{\, AC} \\
     0.030 + 0.0111 e^{0.701 (m_G -12)} \textrm{ mas}& ; \textrm{\, AL}.
     \end{cases}
\end{align}
The uncertainties inside the curly brackets are for a single field-of-view. Since, we are treating successive observations by both fields-of-view as a single pass, we have divided these uncertainties by a factor of $\sqrt{2}$ to characterize the global astrometric position error.

\begin{figure}
	\centering
	\includegraphics[scale=0.8]{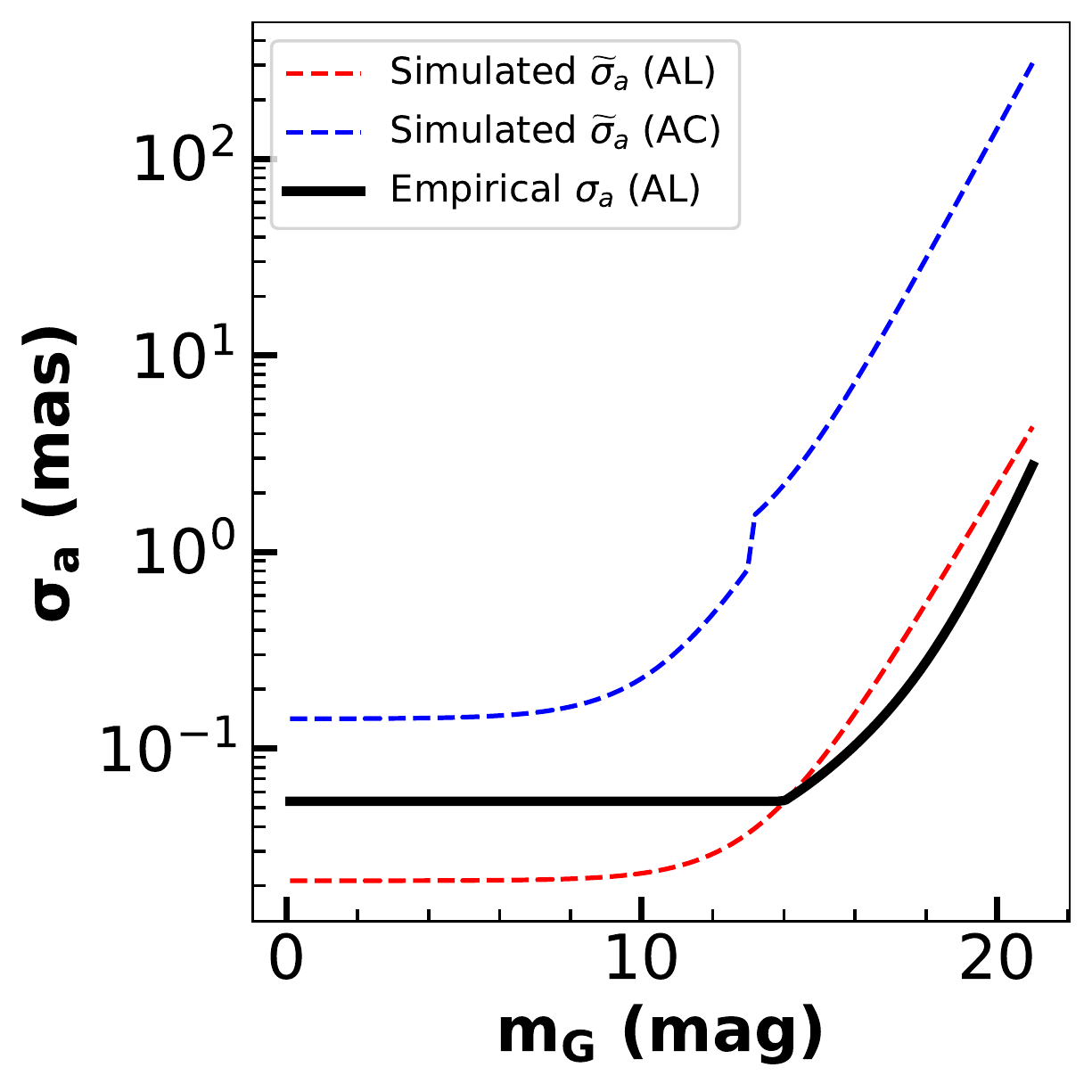}
	\caption{This figure shows the dependence  of single pass uncertainty of the angular position of a star with a given apparent G-band magnitude ($m_G$). The red-dashed and blue-dashed curves show the simulation based uncertainty ($\widetilde{\sigma}_a$)  for both the along (AL) and the across (AC) scanning directions. When computing our results for the expected event rate and event observable distributions, we use the empirically determined uncertainty ($\sigma_a$) along the scanning direction, shown in black.}
	\label{fig:sigma}
\end{figure}

Alternatively, we can use the empirically observed uncertainty along the scanning direction as reported in \Gaia{}~DR2~\cite{2018A&A...616A...2L}. We use a fit to this empirically determined uncertainty using the fitting function of ref.~\cite{2020A&A...640A..83K} given by,

\begin{align}
\label{eq:Asigma}
    \sigma_{a} = & \frac{\sqrt{-1.631 + 680.766 z + 32.732 z^2}\times7.75+100}{\sqrt{N_\textrm{CCD}}} \, \, \mu\textrm{as},
\end{align}
where,
\begin{equation}
    z = 10^{(0.4(\textrm{max}(m_G,14) - 15))}.
\end{equation}
In this formula, we use $N_\textrm{CCD} = 18$ as the number of CCDs in both telescopes together in which each star is imaged along the scanning direction.

We have plotted the simulation based uncertainties (AC and AL) as well as the empirically determined uncertainty (AL) as a function of $m_G$ in fig.~\ref{fig:sigma}. From the figure we can see that all the uncertainties are roughly constant for bright stars with $m_G\lesssim 13$. However, for fainter stars with $m_G \gtrsim 13$ the astrometric uncertainty degrades rapidly with increasing apparent magnitude. We also note that there is a slight difference between the empirical AL uncertainty and the simulated AL uncertainty. In practice, the typical astrometric uncertainty will lie between the uncertainties of the along and across scanning directions.

We will compute the expected exclusion on the PBHs parameter space in sec.~\ref{sec:Results} and the expected event observable distribution in sec.~\ref{sec:lensingrate}, using a fiducial choice of the empirically determined AL uncertainty ($\sigma_a$) as a reference uncertainty. This model gives a roughly constant value of $\sigma_{a} = 0.076$~mas up to $m_G \sim 14$.

\subsection{\Gaia{} eDR3 and stellar catalog}
\label{subsec:GaiaEDR3Catalog}
\Gaia{} has released an early version of its 3rd release of data (eDR3)~\cite{2021A&A...649A...1G} which contains details of sources that it has observed between July 2014 and May 2017 (34 months). This data release does not include time-series information of different passes for individual stars. Rather the \Gaia{} collaboration fits the time-series data to ``astrometric solutions'' which are parametric trajectories for each star which are assumed to only have rectilinear motion relative to the sun.

There are three possible astrometric solutions which are used to fit the trajectories of stars in the catalog. The type of solutions used to fit a particular star's trajectory depends upon the quality of data available for that star~\cite{2012A&A...538A..78L}. In particular, good quality color information is important for correcting chromatic effects which can lead to an offset in the image centroid~\cite{2016A&A...595A...3F}.

For those sources for which color information is of high quality and chromatic effects can be well corrected for, \Gaia{} reports a 5-parameter (5-p) solution. This solution is parameterized by the parallax ($\omega$), right ascension (r.a.)  ($\alpha$), declination (dec.)  ($\delta$), angular velocity along r.a. ($\mu_{\alpha*}$), and angular velocity along dec. ($\mu_\delta$)\footnote{$\mu_{\alpha*}$ is defined as $\mu_{\alpha*} \equiv \frac{d\alpha}{d t}\cos{\delta}$, whereas $\mu_{\delta} \equiv \frac{d \delta}{d t}$.}.

When there is lower quality color information \Gaia{} reports a 6-parameter (6-p) solution, which includes all the parameters of the 5-p solution, plus an additional parameter $\nu_{\textrm{eff}}$. This extra parameter is an effective wavenumber, which allows for chromatic effects to be estimated~\cite{2016A&A...595A...3F}.

Finally, when the data quality is insufficient to obtain a 5-p or a 6-p solution, \Gaia{} reports a 2-parameter (2-p) solution. This solution is parameterized only by the r.a. ($\alpha$) and dec. ($\delta$).

The data released by \Gaia{} in eDR3 only specifies the values of these fitted parameters for each star rather than the full time-series data. The number of sources which are fitted to a given type of solution, and their typical astrometric position uncertainties are shown in tab.~\ref{tab:eDR3content}. Note that these uncertainties are on the fitted $\alpha$, $\delta$, which are obtained for multiple epochs of observation; they do not describe the uncertainty of the astrometric position from a single pass.
 \begin{table}[h]
    	\centering
    	\begin{tabular}{ |p{4cm}||p{3cm}|p{5cm}| }
    		\hline
    		Source type & Number of \newline sources & Uncertainty in position\\
    		\hline
    		Total & 1,811,709,771 & \\
            5-parameter astrometry &  585,416,709 & 0.01-0.02 mas ; $m_G<$15, \newline 0.05 mas ; $m_G$=17,\newline 0.4 mas ; $m_G$=20,\newline 1.0 mas ; $m_G$=21 \\
            6-parameter astrometry & 882,328,109 & 0.02–0.03 mas ; $m_G<$15, \newline 0.08 mas ; $m_G$=17, \newline 0.4 mas ;
            $m_G$=20, \newline 1.0 mas ; $m_G$=21\\
            2-parameter astrometry & 343,964,953 &  1-3 mas \\
    		\hline
    	\end{tabular}
    	\caption{This table shows the number of sources present in the \Gaia{} eDR3 catalog which are brighter than $m_G=21$. It also summarizes the distribution of the sources over 5-parameter, 6-parameter, and 2-parameter solutions. The typical uncertainties on the fitted angular position of the sources in the catalog is also shown. These uncertainties corresponds to the data taken by \Gaia{} over 34 months.}
    	\label{tab:eDR3content}
    \end{table}

According to ref.~\cite{2021A&A...649A...1G}, time-series data will be available from the 4th data release (DR4) onwards. In order to estimate the total number of candidate lensing events expected to be seen by \Gaia{} over $t_\textrm{obs} = 5$ years of data collection, we attempt to build a simple mock catalog of the stars that \Gaia{} will see, \emph{and} which will also have high quality time-series data that can be used to identify potential lensing signals due to AML induced deflections in the apparent trajectories of stars. To this end, we make use of the catalog of stars already seen by \Gaia{} and reported in eDR3. While eDR3 only corresponds to 34 months of collected data, it is unlikely that there will be a significant number of new stars discovered by \Gaia{} which will have such a high quality of data so as to be sensitive to lensing signals. Thus we take the eDR3 catalog to be representative of all the stars (with data quality suitable for precision astrometric analysis) that \Gaia{} will see after 5 years of observation. We further select from among these stars to identify those which are likely to be suitable for detecting an AML signature.

For the purposes of estimating the expected number of lensing candidate events, we need to know the 3D position of each star ($\alpha$, $\delta$, $D_s$), where $D_s$ is the distance to the star, and the apparent $G$-band magnitude $m_G$ of the star. The position is needed to estimate the number of lenses between us and the star and the $G$-band magnitude is needed to estimate the astrometric uncertainty in the position of a star from a single pass. $D_s$ can be inferred from the parallax and thus can only be obtained for those stars for which 5-p or 6-p solutions are available. Thus, we prune our selection of \Gaia{} eDR3 stars to include only those which have either 5-p or 6-p solutions. Also, since for a large number of stars the observed parallaxes have large uncertainties, we use distances from the GeDR3\_dist catalog~\cite{2021AJ....161..147B}. This catalog converts parallax to distance after taking into account priors from a stellar distribution model for the Galaxy\footnote{We are trying to build a model for the stars in the Galaxy, the prior model of the stellar distribution in ref.~\cite{2020PASP..132g4501R} is informed and updated by \Gaia{} eDR3 data. These distances are not used as predicted measurements, but rather we use them in later sections to estimate the AML probabilities and lensing rates.}.

\begin{figure}
    \centering
    \includegraphics[scale=0.45]{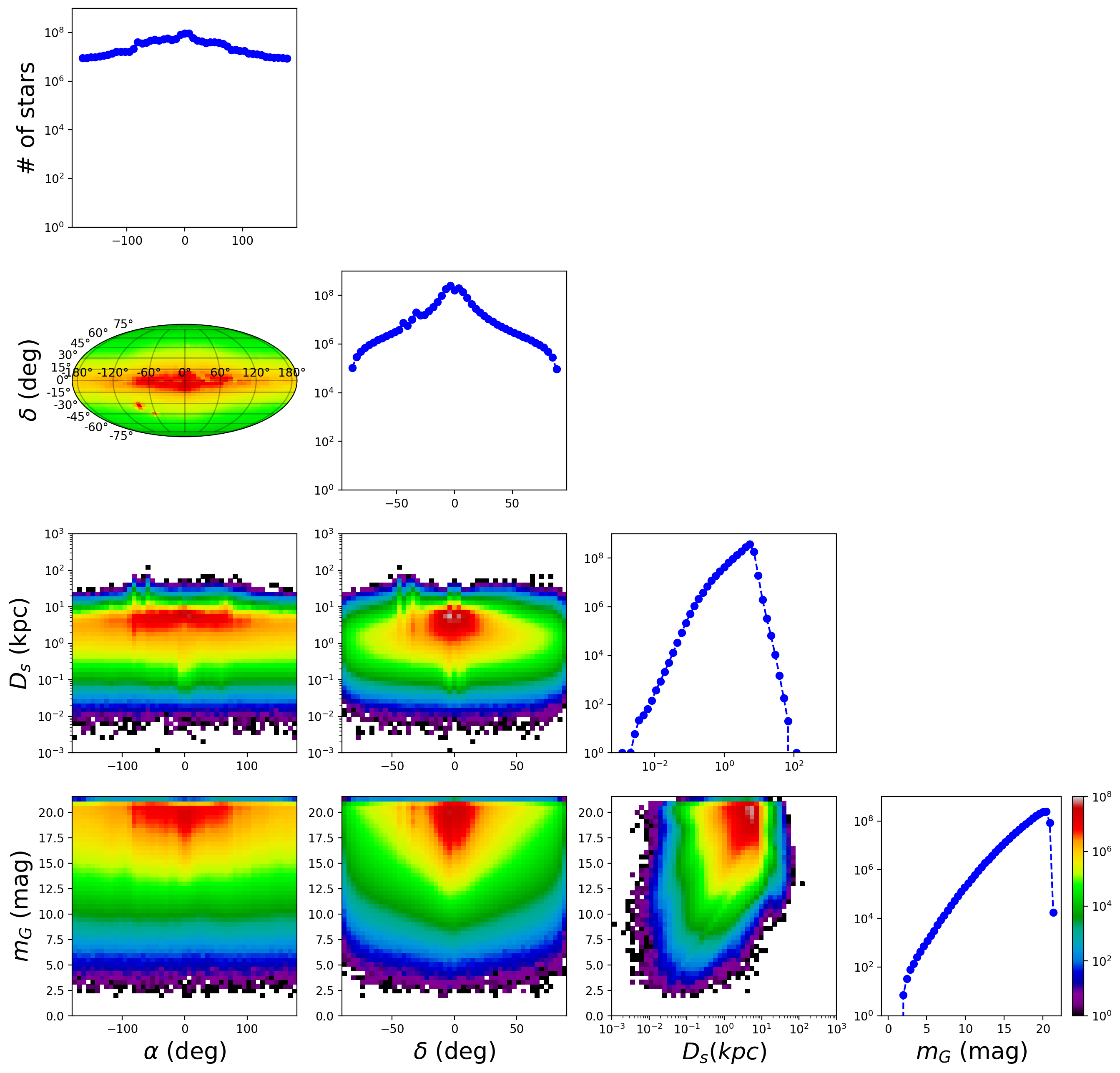}
    \caption{We show the distribution of some properties of the approximately 1.47 billion sources in the \Gaia{} eDR3 catalog with 5-p or 6-p astrometric solutions. We use this stellar distribution as a model for sources that \Gaia{} can track in order to look for signatures of astrometric microlensing due to PBHs. We show the distribution of sources as a function of their Galactic latitude ($\delta$) and longitude ($\alpha$), distance ($D_s$), and apparent G-band magnitude~$m_G$. The distance to each star is taken from the GeDR3\_dist catalog~\cite{2021AJ....161..147B}.
    }
    \label{fig:GeDR3grid50}
\end{figure}

Once we apply the above selection criteria, the total number of stars in our filtered catalog is $\sim1.47$ billion. We have plotted the distribution of these stars as a function of $\alpha$, $\delta$, $D_s$, and $m_G$ in fig.~\ref{fig:GeDR3grid50} (here, and in future sections we will switch conventions and use $\alpha$, $\delta$ to represent Galactic longitude and latitude, respectively). We can see from the figure that most stars are unsurprisingly concentrated near the Galactic center at $\alpha = \delta = 0^\circ$. We can also see from the the $m_G$ distribution of the stars in the filtered catalog (bottom-right panel in the figure), that the number of fainter objects increases up to $m_G=20.7$ and then cuts-off sharply. Although, \Gaia{} may yet discover more faint stars, omitting such stars from our mock catalog will not affect our final lensing event rate prediction since such faint stars are expected to have very large astrometric errors on their positions which will make detecting a lensing signal difficult. This can be seen by comparing the $m_G$ distribution of the catalog (bottom right panel of fig.~\ref{fig:GeDR3grid50}) with fig.~\ref{fig:sigma}, which shows the astrometric positioning uncertainty as a function of $m_G$.

The catalog that we are using contains all point-like sources that \Gaia{} sees. This would include all types of stars such as variable stars, white dwarfs etc., but also luminous non-stellar objects such as brown dwarfs, hot-luminous planets, etc. These objects can also be used as sources to detect AML signatures. The catalog also might contain several extra-galactic objects such as stars in the LMC and SMC, quasars, supernovae, etc. The distance to the extra-galactic objects taken from GeDR3\_dist would be incorrect because the prior model would attempt to place these objects in the Galaxy~\cite{2021AJ....161..147B}. However, since all of the extra-galactic objects only make up a tiny fraction of the catalog, we expect that this will not have a significant effect of our predictions of the AML signal rates\footnote{In ref.~\cite{2021A&A...649A...7G}, the authors attempted to extract LMC and SMC candidate objects in \Gaia{} eDR3 and found only 11,156,431 and 1,728,303, respectively. This is less than 0.1\% of the 1.8 billion objects in the full catalog.}.

In our analysis, we will consider all objects in the \Gaia{} catalog to be point sources (we will refer them as stars) within the Galactic DM halo for purposes of estimating the lensing rate due to PBHs.

%~~~~~~~~~~~~~~~~~~~~~~~~~Section for the lensing rate~~~~~~~~~~~~~~~~~~~~~~~~~~~~~~~~~
\section{Lensing probability}
\label{sec:LensingProb}

In this section we will present analytic formulae for the probability of a particular source to be lensed by a PBH DM candidate. We will first review some basics of AML in sec.~\ref{subsec:AMLbasics}. Then in sec.~\ref{subsec:EventDurationsClassifications}, we will present a formula for the duration of AML signals that are potentially visible to \Gaia{} and we will classify lensing events into one of three types based on their event durations: (i)~short duration lensing events (SDLEs) -- those which have lensing event durations which are shorter than \Gaia{}'s sampling time $t_s = 52.2 \textrm{ days}$, (ii)~intermediate duration lensing events (IDLEs) -- those which have event durations longer than \Gaia{}'s sampling time but shorter than the mission lifetime $t_\textrm{obs} = 5$ years, (iii)~long duration lensing events (LDLEs) -- those which have event duration longer than $t_\textrm{obs}$. Previous literature~\cite{2000ApJ...534..213D}, has primarily focused on LDLE detection by \Gaia{}. As we shall show, SDLEs and IDLEs will also lead to distortions in the apparent trajectories of background stars which possibly could be detectable by \Gaia{}. Because of the short durations of SDLEs, they will have a much smaller detection probability compared to IDLEs and in addition, as we will argue in sec.~\ref{sec:background_stat}, they will be completely dominated by the statistical background and thus will be unimportant for constraining PBHs. On the other hand, the study of IDLE events will be shown to be of utmost importance for detection of relatively lower mass PBHs of mass ($0.1-100 ~M_\odot$). Our treatment of IDLEs is a novel feature of this work. Then, in sec.~\ref{subsec:AMLprob}, we will present analytic formulae for the lensing probabilities in terms of a line-of-sight integral to a source of magnitude $m_G$ with co-ordinates $(D_s,\alpha,\delta)$. In the last two sub-sections we will address some subtleties of our calculation.

\subsection{AML basics}
\label{subsec:AMLbasics}

When a star and PBH lens are nearly aligned, gravitational lensing effects are important. Since, both the stars and PBH lenses have proper motion in the sky, lensing signals will be transient phenomena. Before discussing the transient behaviour and durations of such lensing events, we will first discuss the simple case of a static source and a static lens, relative to us, in the sky. A more detailed version of the discussion that follows can be found in refs.~\cite{1992grle.book.....S, 1996astro.ph..6001N, 1998LRR.....1...12W}.

We take the source to be located at a distance $D_s$ from us and the lens is assumed to be at a distance $D_l<D_s$, and is assumed to have mass $M$. We will denote the angular separation between the source and lens as $\theta$. Gravitational lensing results in magnified multiple images of a source due to the presence of a massive lensing object in the foreground. We will focus on the case of point like source and a point like lens. Lensing phenomena in such a case are characterized by an Einstein ring defined as a circle centered at the lens position on the celestial sphere with an angular size given by the Einstein angle,
\begin{align}
\label{theta_E}
\theta_E & = \sqrt{\frac{4G M }{c^2}\frac{D_s-D_l}{D_l D_s}}, \nonumber \\
& \approx 2.85 \text{ mas }\sqrt{ \frac{M}{1 ~M_{\odot}}\frac{1 \textrm{ kpc}}{D_s}\frac{D_s - D_l}{D_l}}.
\end{align}
Gravitational lensing results in two distinct images of the source which are located on either side of the true source position, along the line joining the source and the lens on the celestial sphere. The angular separation between each of these images and the lens position is given by,
\begin{equation}
\theta_I^\pm = \frac{\theta_E}{2}\left(u \pm \sqrt{u^2 +4}\right),
\end{equation}
respectively, where $u \equiv \theta / \theta_E$ is the source position relative to the lens, scaled by the Einstein angle. Both these images will have different magnification in general, with the magnifications given by the following expressions,
\begin{equation}
m_I^\pm =  \frac{u^2 + 2}{2u \sqrt{u^2 + 4}} \pm \frac{1}{2}.
\end{equation}

It is almost always the case that for the lens masses that we will be considering, the two images will not be resolvable by \Gaia{}. In such a case, we would only detect a single image located at the ``center-of-light'' (centroid) of the two images\footnote{Center-of-light is defined as magnification-weighted average position of the two images.}.

In general this image is magnified and shifted relative to what we would have seen if there were no lens. The relative magnification of this image, as compared to the unlensed case, is given by,
 \begin{equation}
 \label{eq:CentroidMag}
    m = \frac{u^2+2}{u\sqrt{u^2+4}}.
 \end{equation}

We denote the angular shift in the centroid away from the true source position on the celestial sphere as $\delta \theta = \theta_E \delta u $, where $\delta u$ is the angular shift in units of $\theta_E$ and is given by~\cite{1995ApJ...453...37W},
\begin{equation}
    \label{eq:CentroidPos}
    \delta u = \frac{u}{u^2 + 2}.
\end{equation}
This shift is along the line joining the lens to the true position of the source, and it is directed away from the lens.

\begin{figure}
    \centering
    \includegraphics[scale=0.8]{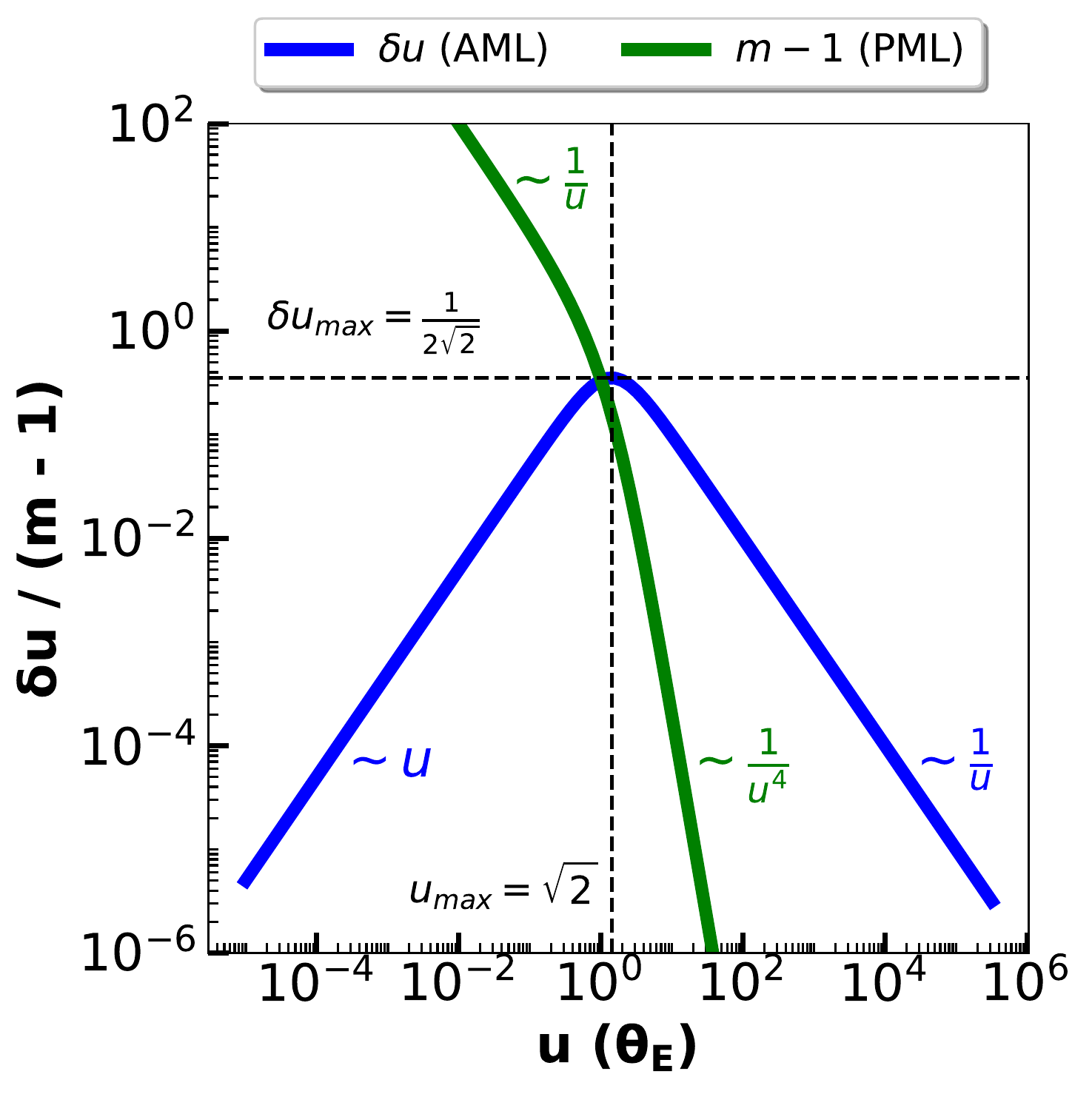}
    \caption{ This figure shows the AML observable, i.e. the shift in angular position $\delta u$, and the PML observable, i.e. the change in relative magnification $m-1$, as a function of $u$, the angular separation between the star and the PBH in units of the Einstein angle, $\theta_E$. The AML observable falls off as $\frac{1}{u}$ for large $u$, whereas the PML observable falls of as $\frac{1}{u^4}$ for large $u$. Also note that unlike the PML observable, the AML observable decreases for small separation between the source and the lens. The AML observable peaks for an angular separation $\delta \theta = \sqrt{2} \theta_E$.}
    \label{fig:delta-u}
\end{figure}

Photometric microlensing attempts to detect the magnification change, whereas astrometric microlensing attempts to detect the angular shift of the image. In fig.~\ref{fig:delta-u}, we show the astrometric shift, $\delta u$ and the change in photometric magnification, $m-1$ as a function of the scaled relative angle between the true position of the source and the lens, $u$.

Both the change in magnification and the angular shift are small for large separation between the source and the lens -- typically for a separation larger than a few $\theta_E$. However, as we can see from the figure and the formulas above, the change in magnification, $m-1$, falls off as $1 / u^4$, whereas the shift in angular position, $\delta u$, falls off as $1 / u$. This implies that, for large separation between the source and the lens, astrometric microlensing can be more sensitive to the presence of a lens. In practice, the relative sensitivity of these two microlensing techniques also depends on the instrumental sensitivity to changes in magnification versus sensitivity to astrometric position measurements. In the rest of this section, we will focus only on the astrometric microlensing technique.

\subsection{Event durations and classifications of AML signals}
\label{subsec:EventDurationsClassifications}
Now we consider the effect of relative motion between the star and the lens on the apparent trajectory of the star. The relative motion of the star and the lens will change their relative separation parameter $u$. This in turn will lead to a change in the astrometric shift $\delta \theta$ of the apparent source position over time.

For simplicity, we will consider the case where only the source is in motion with a tangential speed $v_s$ relative to us, and the lens is assumed to be static. We will also ignore the effects of parallax due to the earth's motion~(similar to refs.~\cite{2000ApJ...534..213D, Belokurov:2001vh}). We will comment on how the computations of this subsection can be generalized to scenarios where both the lens and the source are moving in sec.~\ref{subsec:lensmotion}. We will also discuss the effect of the second assumption (the neglect of parallax) on our analysis in sec.~\ref{subsec:parallax}.

We define the lens plane as a plane perpendicular to our line-of-sight which contains the lens. The source's position projected on to this plane moves with an angular speed $\mu \approx v_s/D_s$, where $D_s$ is the distance to the source. We will then define an effective speed parameter $v\equiv v_s\frac{D_l}{D_s}$ such that $\mu = v/D_l$, where $D_l$ is the distance from us to the lens. A further discussion and interpretation of the parameter $v$ as a relative separation velocity between the source and the lens will be discussed in sec.~\ref{subsec:lensmotion}. The parameter $v$ will crucially determine the duration of lensing events as well as their probabilities.

The astrometric lensing effect can be treated as the instantaneous displacement of the apparent source position relative to the true position of the source, as the source moves across the sky. Up to an overall rotation of the source-lens configuration, a general trajectory of the source in the presence of a foreground lens can then be characterized by the speed $v$ and the impact parameter $u_0$. The impact parameter is defined by the angular separation at the point of closest approach between the source and the lens, when the source's position is projected onto the lens plane.

\begin{figure}
	\centering
	\includegraphics[scale=0.8]{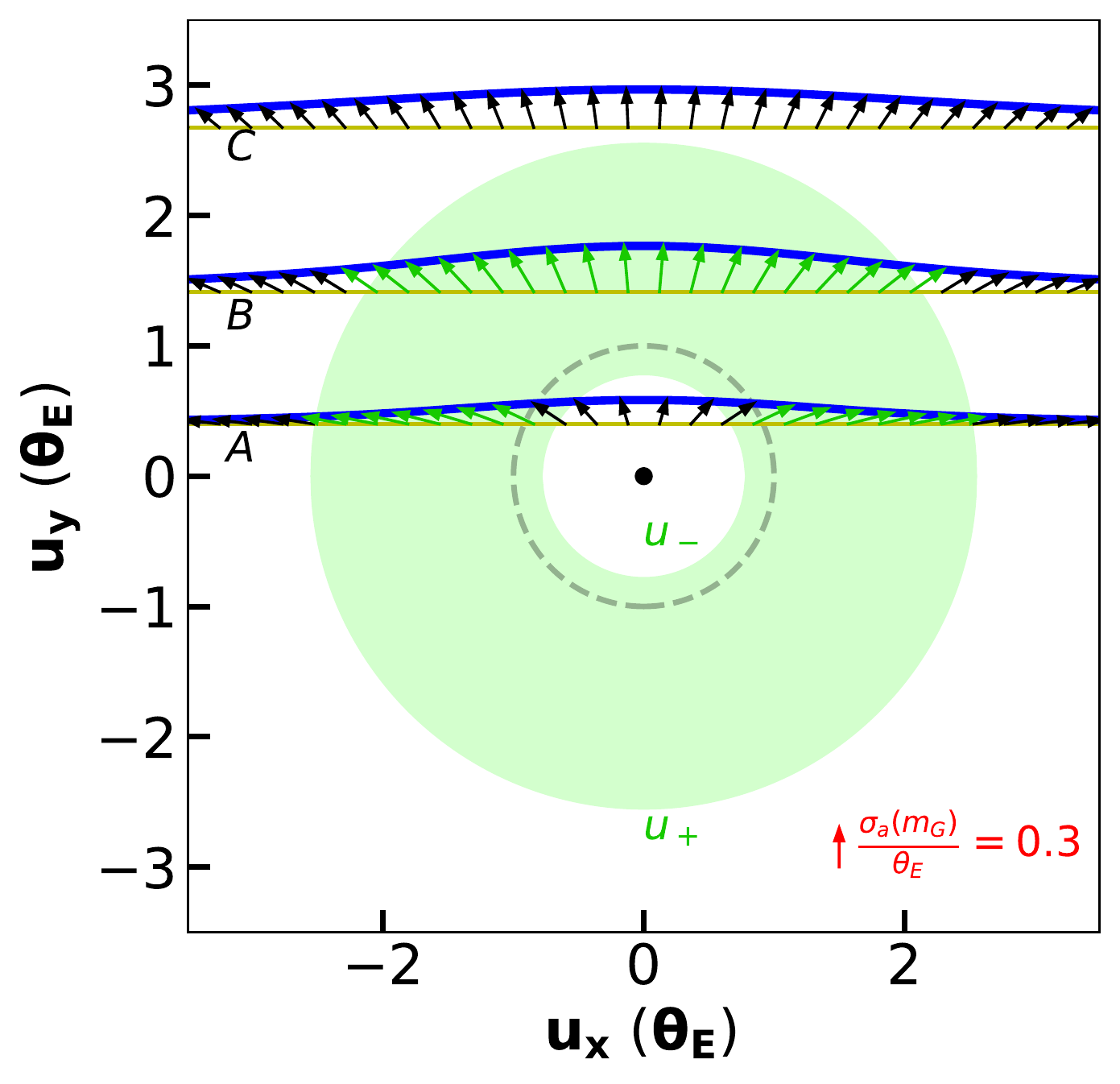}
	\caption{The figure shows three rectilinear trajectories of stars labelled $A$, $B$, $C$ (yellow horizontal lines), and the distortion in the apparent trajectories (blue curves) due to a PBH lens in the foreground (black dot at the center). The Einstein ring around the PBH is shown by the grey dashed circle. The arrows from the yellow to the blue curves indicate the size and direction of the astrometric shift of the apparent position, away from the true position, due to the lensing effect of the PBH. The direction of the shift is always radially away from the PBH. The magnitude of the shift is above \Gaia{}'s sensitivity threshold (reference arrow $\frac{\sigma_a}{\theta_E}$ shown in the bottom right) in the green annular region around the lens. The inner radius of this annulus is $u_-$, and the outer radius is $u_+$. The image is color coded so that the deviation vectors are in green if the shift is above threshold, and they are shown in black otherwise. Trajectories $A$ and $B$ have some portion which lies in the annular region, however, trajectory $C$ lies wholly outside the annular region and will not give rise to a detectable AML signal at \Gaia{}. The event durations for trajectories $A$ and $B$ are given by the time that the star spends in the annular region. We classify AML events based on their average lensing event durations into either SDLEs, IDLEs, or LDLEs.}
	\label{fig:EventDuration}
\end{figure}

In fig.~\ref{fig:EventDuration}, we show in yellow, several rectilinear trajectories, with different impact parameters for a background star. Here the lens, indicated by the black dot, is located at the center of the figure. In this figure, we are showing relative angular positions $u_x, u_y$ on the celestial sphere, scaled to the Einstein angle $\theta_E$. The Einstein ring ($u = 1$) is indicated by a grey dashed circle. At every position the star undergoes an AML effect which leads to a shift in its apparent position relative to its true position. This deflection is along the line joining the lens to the source, and directed away from the lens. These deflections from the true trajectories are denoted with arrows in the figure. The locus of all the arrow-tips, shown in blue, indicates the apparent lensed trajectory that we would see. The effect of lensing is thus to create a distortion of the apparent trajectory away from the true simple rectilinear motion.

By examining the observed trajectory of a star and looking for deviations from rectilinear motion, we can detect signatures of PBH lenses. Next we discuss whether such deviations can be detected by \Gaia{}.

\vspace{3mm}

\Gaia{}'s sensitivity to AML distortion in the trajectory will depend on three factors:
\begin{enumerate}
    \item The uncertainty in the position measurement of a star in a given pass.
    \item The rate at which \Gaia{} makes observations of a particular star's trajectory (cadence).
    \item \Gaia{}'s total observation time $t_\textrm{obs}$.
\end{enumerate}

As discussed in sec.~\ref{sec:Gaia}, \Gaia{} scans a particular region of the sky after every $t_s$, where $t_s \sim 52.2$ days is the sampling time. In a given sampling, the uncertainty on the source position measurement is given by $\sigma_a(m_G)$ and depends on the G-band magnitude $m_G$ of the star, see sec.~\ref{subsec:GaiaDetProperties}. Thus, a minimum criteria for \Gaia{} to be sensitive to the deviation in the star's lensed trajectory relative to its true trajectory, is that the astrometric shift $\delta \theta = \theta_E \delta u$ should exceed the astrometric resolution $\sigma_a(m_G)$. It is easy to show from eq.~\ref{eq:CentroidPos} that this condition can be satisfied for an angular separation between star and lens (expressed in $\theta_E$ units) which lies between $u_+$ and $u_-$ where,
\begin{equation}
\label{eq:uplusminus}
    u_{\pm} = \frac{1}{2} \left(\frac{\theta_E}{\sigma_a(m_G)} \pm \sqrt{\frac{\theta_E^2}{\sigma_a(m_G)^2} - 8}\right).
\end{equation}
In fig.~\ref{fig:EventDuration}, we also show in green, an annular region around the lens, where the inner radius of the annulus is $u_-$ and the outer radius is $u_+$. A star passing through this annular region will experience a deflection in its trajectory which is larger than \Gaia{}'s single pass astrometric resolution.

In the figure, we have shown a representative measure of the single pass astrometric resolution $\sigma_a(m_G)/\theta_E = 0.3$, as a red arrow annotation at the bottom right. The deflection arrows along each trajectory are color coded such that the arrows are show in black if the deflection is smaller than this threshold, and the arrows are shown in green if they are larger than this threshold. For the portion of the trajectory that lies within the annular region, the deflection arrows are green, indicating that the deflection in the trajectory is above \Gaia{}'s detection threshold.

In the figure, trajectory $A$ enters the green annular region and thus creates an above threshold deflection, after which it enters the inner annular region and the deflection falls below threshold. Then it once again enters the annular region and the deflection is above threshold, before finally exiting the annular region, and the deflection drops below threshold once again. Trajectory $B$ enters the annular region and creates an above threshold deflection, but it does not enter the inner annular region before exiting the annulus. Finally, trajectories with a large impact parameter ($u_0 > u_+$), such as trajectory $C$, will lie wholly outside the annulus and will always have deflections below threshold. Such trajectories will always yield undetectable lensing signatures.
For $\theta_E<2\sqrt{2}\sigma_a(m_G)$, the annulus shrinks to zero size, and the lensing signal would be too weak to be detectable for any impact parameter. Thus, for $\theta_E>2\sqrt{2}\sigma_a(m_G)$, and trajectories with impact parameter $u_0 < u_+$, we will have AML deflections above \Gaia{}'s threshold sensitivity, and thus, a potentially detectable lensing signature.

For a source moving with angular speed $\mu$, we can define a lensing duration $t_e$ for a given trajectory, during which the deflection of the apparent trajectory from the true trajectory is above \Gaia{}'s sensitivity threshold, i.e. $\delta \theta > \sigma_a(m_G)$\footnote{For photometric microlensing, we can also define event duration for which the $m-1 > \sigma_p(m_G)$, where $\sigma_p(m_G)$ is the uncertainty in the apparent magnitude $m_G$. Typically we expect that the duration of the photometric event would be much smaller than the astrometric event. This is because photometric signal ($m-1\sim 1/u^4$) decays much faster than astrometric signal ($\delta \theta \sim 1/u$) with the angular separation $u$. However, an exact comparison depends on the photometric and astrometric resolutions.}. Thus, $t_e$ is just the time that would be spent by the star in the annular region. Note that this time $t_e$ need not be something that \Gaia{} actually observes, rather it should be thought of as a parameter associated with a source-lens configuration that will help us quantify whether the lensing signature is detectable by \Gaia{}.

We can divide lensing events into three categories based on the value of $t_e$,
\begin{itemize}
    \item \textbf{Short duration lensing events (SDLEs)} -- those with $ t_e  < t_s$. In this case, these events will most likely be missed by \Gaia{}, as the above threshold deflection is most likely to occur in between \Gaia{}'s samplings of the region of the sky containing the source and the lens. However, it is possible that a single sample of the trajectory may show a large deviation due to an SDLE which coincidentally occurs when \Gaia{} happens to be observing a star while it is being lensed.
    \item \textbf{Intermediate duration lensing events (IDLEs)} -- those with $t_s< t_e  < t_{\text{obs}}$. In this case, most of the trajectory that \Gaia{} observes will not suffer significant distortion due to lensing. However, there will be at least a few sample observations seen along the trajectory that show a lensing deflection $\delta \theta > \sigma_a(m_G)$. Such a trajectory can be easily distinguished from rectilinear motion because of the blip/distortion in the apparent trajectory.
    \item \textbf{Long duration lensing events (LDLEs)} --  those with $  t_{\text{obs}} <  t_e $. For these trajectories, almost the entire trajectory suffers from a significant distortion due to lensing, and in the extreme limit of large event duration, \Gaia{} will only see a fraction of the lensed trajectory. Counter-intuitively, this can make detection of such signals harder than the previous case since we will not have a reference for rectilinear motion against which we can detect a lensing signature.
\end{itemize}

In summary, detectable lensing events at \Gaia{} will require the following conditions to be met for the source-lens configuration.
\begin{enumerate}
\item First, the annular region of source positions around the lens, in which the deflection of the star's apparent position is above threshold, should exist. This condition requires $\theta_E > 2\sqrt{2} \sigma_a(m_G)$.
\item Second, a portion of the star's trajectory must pass through this annular region.
\item The lensing event duration $t_e$, which is the time spent by the star in the annular region will determine the type of event. For $t_e < t_s$ and assuming that the above two conditions are satisfied, we will get an SDLE type of event. However, the probability of detection of such an event will be suppressed by the coincidence probability $t_e/t_s$ of sampling a star while such a lensing event is in progress. For $t_s< t_e < t_\textrm{obs}$, we will get an IDLE type of event, which we will assume will be detected by \Gaia{} as long as the above two criteria are satisfied. For $t_e>t_\textrm{obs}$, we will have LDLE type events if both of the above criteria are satisfied. In addition, we will also assume for LDLE events a threshold criteria for detection by \Gaia{} by requiring that the \textit{difference} between the deflections at the initial and final epochs of observation differ by an angle greater than $\sigma_a(m_G)$ (see ref.~\cite{2000ApJ...534..213D}).
\end{enumerate}

Assuming that the annular region exists, and a star and lens come close enough to each other that the star passes through this region, we would like to estimate what the typical event durations would be. We compute the event duration averaged over viable impact parameters which we denote as $\langle t_e \rangle$.  For a given impact parameter, the event duration is $t_e = l(u_0)/\mu$, where $l(u_0)$ is the (angular) length of the portion of the trajectory which lies within the annulus. Integrating this time over $u_0$ and dividing by the allowed range of $u_0$ gives the average event duration $\langle t_e \rangle$. The allowed range of impact parameters for which the star will pass through the annular region around the lens is $0<u_0<u_+$. The integration of $l(u_0)$ over $u_0$ therefore gives the area of the upper half of the annulus. Thus we have,

\begin{align}
\label{eq:eventduration}
\langle t_e \rangle = & \frac{\pi \theta_E (u_+^2 -u_-^2)}{2 u_+ \mu}, \nonumber\\
                    = & \frac{\pi D_l \theta_E}{ v} \frac{\frac{\theta_E}{\sigma_a(m_G)} \sqrt{\frac{\theta_E^2}{\sigma_a(m_G)^2} - 8}}{\frac{\theta_E}{\sigma_a(m_G)} + \sqrt{\frac{\theta_E^2}{\sigma_a(m_G)^2} - 8}},
\end{align}
where the quantity under the square root is positive for detectable lensing events.  For a $10~M_{\odot}$ lens, with source and lens at a distance of $\mathcal{O}(1) \text{ kpc}$ from us, and an astrometric precision of 1~mas, we would find that the average event duration is $\sim 1-2 $~\text{ years}, giving us an IDLE type event. For larger masses for the PBH, or better astrometric precision (brighter sources), or larger distances to the source, we can get event durations  longer than the \Gaia{} mission time $t_{\text{obs}}$, i.e. we would get LDLE type events.

\subsection{AML event probability due to PBH DM}
\label{subsec:AMLprob}

Given an assumption of the PBH parameters $(f, M)$, we will now derive an expression for the probability $P_\textrm{star}$ for a source with apparent magnitude $m_G$ and Galactic co-ordinates $(D_s, \alpha, \delta)$ to undergo a lensing event which is detectable by \Gaia{}. This probability will be the sum of the probabilities for a source to yield either an SDLE, IDLE, or an LDLE type event.

We split this calculation into two parts, first we will calculate the conditional probability of lensing of such a star $p_\textrm{lensing}^c$, assuming that there is a lens located at a distance $D_l$ between us and the source, such that the lens is localized within a solid angle $\Delta \Omega$. Then we will weight this conditional probability by the probability that such a lens is actually present, assuming that the lenses are in the form of PBHs of mass $M$, which make up a fraction $f$ of the DM in the Galactic halo.

We will first compute $p_\textrm{lensing}^c$. Let us consider a patch of sky that subtends a solid angle~$\Delta \Omega$. We assume that this patch contains a source located at a distance $D_s$ from us, and a lens located at a distance $D_l$ from us. As the source moves across the sky over a time $t_{\text{obs}}$, the source traverses an angular distance on the sky $\mu t_{\text{obs}}$, where $\mu = v/D_l$ is the proper speed of the source. We assume that $\Delta \Omega$ is sufficiently large such that the entire trajectory of the star is well contained within this patch of the sky. If the lens lies within an impact parameter $u_+ \theta_E$ on either side of the star's trajectory, a portion of the trajectory will have an AML deflection which is larger than \Gaia{}'s threshold resolution $\sigma_a(m_G)$. However, the impact parameter $u_0$, and hence the event duration $t_e$, will depend upon the angular orientation of the source's velocity vector. We take our formula in eq.~\ref{eq:eventduration} for $\langle t_e \rangle$, averaged over impact parameters, to represent the typical event duration averaged over orientations of the source velocity vector.

For IDLEs ($t_s<\langle t_e \rangle<t_\textrm{obs}$), the bulk of the trajectory is sampled while the star is not undergoing any lensing signal and therefore the apparent motion appears rectilinear. However, for a small portion of the trajectory, during which it is sampled at least a few times, the trajectory will show a clear deviation from rectilinear motion, characteristic of an IDLE. In this case we assign $p_\textrm{lensing}^c = \frac{\delta \omega_{\textrm{IDLE}} }{\Delta \Omega}$, where $\delta \omega_{\textrm{IDLE}} =\mu t_\textrm{obs} \times 2 u_+ \theta_E $ is the angular area of an imaginary rectangle swept out by the source as it moves across the sky, of width $2 u_+ \theta_E$. If the lens lies within the rectangle, it would lead to an IDLE type event.

For SDLEs, $p_\textrm{lensing}^c$ will be same as that for IDLEs, except for an additional coincidence probability suppression factor of $\langle t_e\rangle/t_s$.

For LDLEs ($t_\textrm{obs} < \langle t_e \rangle$), although the event duration is large, and the trajectory will be sampled frequently during a period in which the deflection of the source's apparent position from its true position is above the \Gaia{} threshold, there is no reference rectilinear part of the apparent trajectory against which we can compare this deviation. In this case, a possible requirement for the threshold for detection could be to demand that the difference between the deflections at the start and at the end of the observational period is larger than $\sigma_a(m_G)$. In ref.~\cite{2000ApJ...534..213D}, the authors made exactly this assumption and computed $p_\textrm{lensing}^c = \frac{\delta \omega_{\textrm{LDLE}}}{\Delta \Omega}$, where $\delta \omega_{\textrm{LDLE}} = \mu t_\textrm{obs}\times 2 \sqrt{\frac{ v t_{\text{obs} } }{D_l} \frac{1}{\sigma_a(m_G)}}\theta_E $.

In summary we find,
\begin{equation}
\label{eq:plensing}
    p_\textrm{lensing}^c = \begin{cases}
         \frac{2}{\Delta \Omega} \left(\frac{\theta_E}{\sigma_a(m_G)} + \sqrt{\frac{\theta_E^2}{\sigma_a(m_G)^2} - 8}\right) \theta_E \mu t_{\textrm{obs}} \times \frac{\langle t_e\rangle}{t_s} & ; \,  \langle t_e \rangle < t_s , \\
        \frac{2}{\Delta \Omega} \left(\frac{\theta_E}{\sigma_a(m_G)} + \sqrt{\frac{\theta_E^2}{\sigma_a(m_G)^2} - 8}\right) \theta_E \mu t_{\textrm{obs}} & ; \, t_s < \langle t_e \rangle < t_{\textrm{obs}}, \\
        \frac{2}{\Delta \Omega}  \sqrt{\frac{t_{\text{obs} } v}{D_l} \frac{1}{\sigma_a(m_G)}} \theta_E \mu t_{\textrm{obs}} & ; \,  t_{\textrm{obs}} < \langle t_e \rangle.
    \end{cases}
\end{equation}
This computation of $p_\textrm{lensing}^c$ is valid only when the astrometric deflection is above threshold. To be explicit, we should multiply the right-hand-side of the above formula by the step function $\Theta(\theta_E - 2\sqrt{2}\sigma_a(m_G))$, in order to ensure this condition. In computing this probability above, we have averaged over an assumed uniform distribution of the star and the lens positions within the angular window of size $\Delta \Omega$, and we have also averaged over different orientations of the source velocity vector.

\vspace{3mm}

Now, for a particular star located at a given ($D_s, \alpha, \delta$), we can estimate the probability $P_\textrm{star}$ of such a star undergoing a microlensing event caused by PBHs of mass $M$, that make up a fraction $f$ of the Galactic DM halo. We make the assumption that the probability of lensing of a particular star is small and the lens distribution is diffuse enough such that the apparent trajectory of the star is only affected by a single lens at a time. We call this the unilensing assumption.

$P_\textrm{star}$ can then be computed by first multiplying the conditional probability $p_\textrm{lensing}^c$, weighted by the Poissonian probability of having a single lens in the angular window $\Delta \Omega$ located at a distance between $D_l$ and $D_l + \Delta D_l$, which lies between us and the source. The Poissonian probability can just be taken to be the average number of lenses in the aforementioned region. We then sum this over all distances $D_l$ between us and the source.

Thus we obtain,
\begin{align}
\label{eq:P}
    P_\textrm{star} = & \int_{0}^{D_s} D_l^2 dD_l \Delta \Omega \frac{f}{M}\rho_{\textrm{DM}}(D_l,\alpha,\delta) p_\textrm{lensing}^c,
\end{align}
where we assume that the DM density is given by a standard spherically-symmetric NFW profile~\cite{NFWparameter} about the Galactic center,
\begin{equation}
    \rho_{\textrm{DM}} = \frac{\rho_0}{\frac{r}{r_s}\left(1+ \frac{r}{r_s}\right)^2},
\end{equation}
where $\rho_0 =1.06\times10^7 ~M_{\odot}/\textrm{kpc}^3 $ and $r_s = 12.5 \textrm{ kpc}$ is the DM scale radius. The distance $r$ of a lens from the center of the Galaxy can be written in terms of of the distance of the lens from the earth's position $D_l$, as $r = \sqrt{D_l^2 + r_e^2 - 2 D_l r_e \cos{\delta} \cos{\alpha}}$. Here, $r_e = 8.5 \textrm{ kpc}$ is the distance between the earth and the Galactic center, and $\alpha, \delta$ give the position of the star in Galactic co-ordinates. Note that the $\Delta \Omega$ factor cancels out with the factor $1/\Delta \Omega$ in $p_\textrm{lensing}^c$. Thus, we are left with a single line-of-sight integral to the star in order to compute $P_\textrm{star}$.

For a given hypothesis of the PBH mass $M$ and density fraction $f$, the probability $P_\textrm{star}$ is a function of (i) the location $(D_s,\alpha, \delta)$, (ii) the separation velocity $v$, and (iii) the apparent magnitude $m_G$ (through the astrometric resolution $\sigma_a(m_G)$), of the star. The line-of-sight integral can be numerically evaluated for a given star in the \Gaia{} catalog, under a specific hypothesis for the PBH parameters. Care must be taken to evaluate $\langle t_e \rangle$ to check which case of lensing applies, and care must also be taken to check the value of the step function argument, to ensure that the lensing signal is above threshold. Both these conditions depend on the value of $D_l$ and this complicates the numerical integration of the expression in eq.~\ref{eq:P}.

\vspace{3mm}

To minimize the computational time, instead of evaluating $P_\textrm{star}$ individually for each star in the \Gaia{} catalog, we tabulated its values. First, we fixed the values of PBH parameters $(f,M)$. We assumed a fixed relative separation velocity $v = 200 \textrm{ km/sec}$\footnote{See sec.~\ref{subsec:lensmotion} for a discussion on the implications of this assumption.}. Then, we numerically evaluated the expression in eq.~\ref{eq:P} on a grid of $(D_s,\alpha,\delta,m_G)$ values. We show the range of values of these parameters that we have scanned over in tab.~\ref{tab:GeDR3Coarse-Grained}.

We then repeated our tabulation of $P_\textrm{star}$ for different assumptions of the PBH parameters. The dependence on $M$ is complicated, because $\theta_E$ depends on $M$, and $P_\textrm{star}$ depends on $\theta_E$ in a non-trivial way. However, the dependence of the probability on $f$ can be found by just a trivial overall rescaling.

\begin{table}[h]
    	\centering
    	\begin{tabular}{ |p{3cm}||p{2.5cm}|p{2.5cm}| p{1.5cm}| }
    		\hline
    		Parameters& min value &max value& grid size\\
    		\hline
    		$\alpha$ & -180 deg & 180 deg&50\\
    		$\delta$ & -90 deg & 90 deg&100\\
    		%$D_s$& 0 kpc& 333 kpc&50\\
    		$D_s$& 0.5 kpc& 20 kpc&100\\
    		%$m_G$ & 0.3 & 19.7&50\\
    		$m_G$ & 10 & 21.6&50\\
    		\hline
    	\end{tabular}
    	\caption{Rather than evaluate $P_\textrm{star}$ individually for each star in the \Gaia{} catalog, we tabulate $P_\textrm{star}$ on a grid of ($D_s,~\alpha,~\delta,~m_G$) values. This table show the minimum and maximum values of each parameter, and the grid size gives us the number of samples that we take for each parameter, i.e. we are evaluating $P_\textrm{star}$ on a $50\times100\times100\times50$ grid. In order to evaluate $P_\textrm{star}$ for a given star in the \Gaia{} catalog, we simply read off the value from the nearest point on our grid.}
    	\label{tab:GeDR3Coarse-Grained}
    \end{table}

\begin{figure}
    \centering
    \includegraphics[scale=0.45]{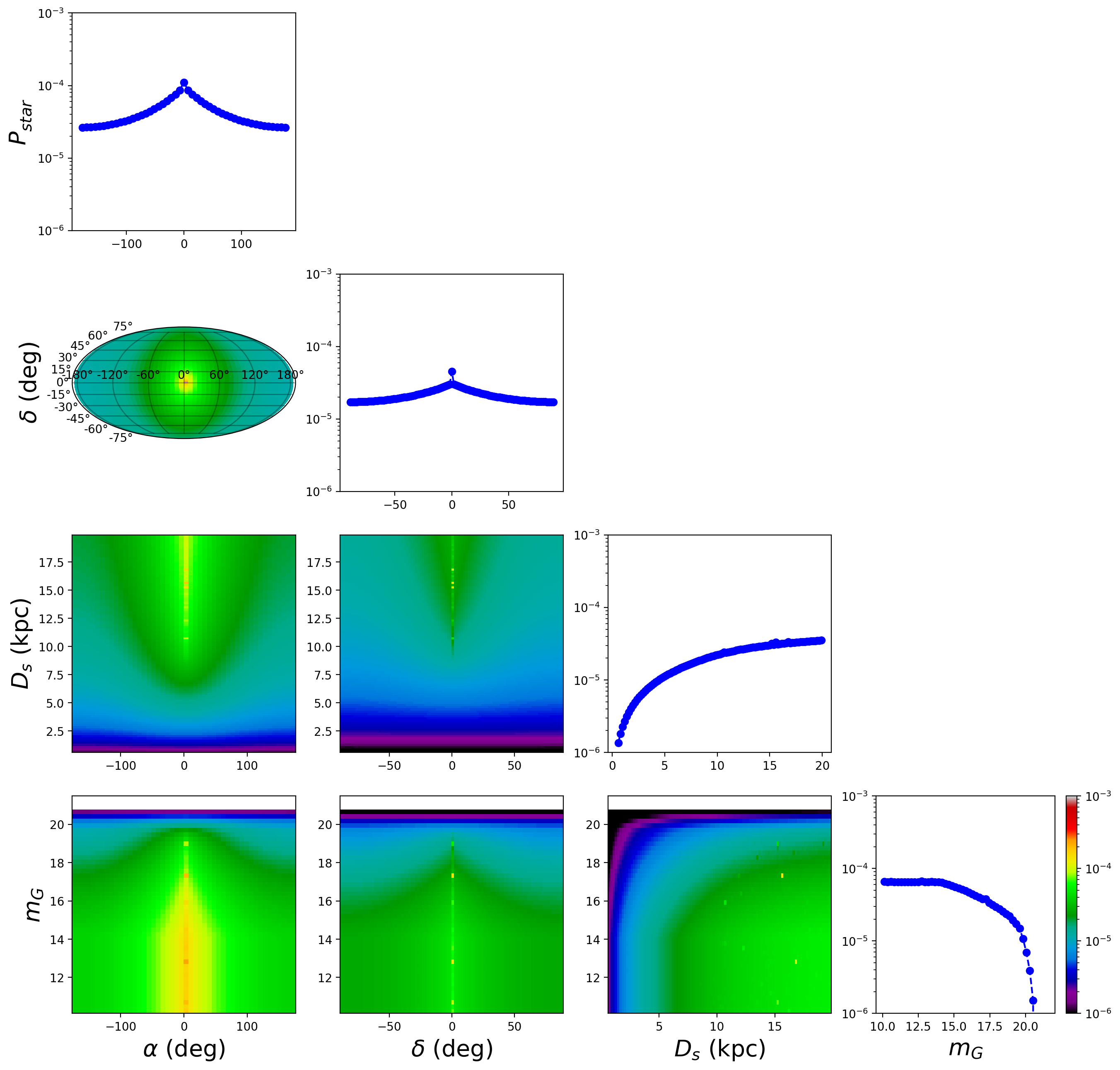}
    \caption{\label{fig:LensingEvents} The probability $P_\textrm{star}$ for a given star located at $(D_s,\alpha,\delta)$ and with apparent magnitude $m_G$ to undergo an astrometric microlensing event which will be detectable by \Gaia{} over an observational time $t_\textrm{obs} = 5~\textrm{years}$. This figure shows the probability for specific PBH parameters $(f,M)$ = $(1,14~M_\odot)$. The plot depicts $P_\textrm{star}$ as a function of any two parameters among $(D_s,\alpha,\delta,m_G)$, while averaging uniformly over the other two parameters over their full range (see tab.~\ref{tab:GeDR3Coarse-Grained}). Also shown is the dependence of $P_\textrm{star}$ on any one parameter, while averaging over the other three parameters. An intuitive explanation for the dependence of this function on each of the parameters is given in the text. In sec.~\ref{sec:Results}, we will weight the stars in the \Gaia{} catalog by this probability distribution to obtain the total number of lensing events that \Gaia{} is expected to detect.}
\end{figure}

We show the results of our numerical evaluation of $P_\textrm{star}(D_s,\alpha,\delta,m_G)$ for $f = 1$ and $M=14 ~M_\odot$ in fig.~\ref{fig:LensingEvents}\footnote{Our plot actually shows $P_\textrm{star}$ only taking into account IDLE and LDLE type events. SDLE type events have a suppressed probability distribution and will in any case have a small contribution to $P_\textrm{star}$ for this PBH mass.}. To visualize this probability we have plotted its distribution as a function of two parameters at a time, while uniformly averaging over the other parameters. We have also shown the 1D distributions of this probability as function of each of the individual parameters.

From the figure, we can see several interesting features. First, the probability of lensing of a star is largest in the direction of the Galactic center $\alpha = 0^\circ$, $\delta = 0^\circ$. This is expected because of the large DM density, and hence large number of potential lenses in that direction. Second, we see that $P_\textrm{star}$ increases as a function of $D_s$ before eventually saturating beyond the NFW scale radius. This saturation is expected, as there is no further significant gain in the number of lenses along any line-of-sight at large Galactic radii. Finally, we see that $P_\textrm{star}$ is almost constant as a function of $m_G$ and then decreases sharply for $m_G\gtrsim15$. This behaviour can be attributed to \Gaia{}'s astrometric position measurement uncertainty which becomes large for $m_G>14$, see eq.~\ref{eq:Asigma}.

We also note here that it is possible to define $P_\textrm{star}^\textrm{SDLE}$, $P_\textrm{star}^\textrm{IDLE}$, and $P_\textrm{star}^\textrm{LDLE}$ as the probabilities for a given star to undergo either an SDLE, IDLE, or LDLE, respectively. These can be computed by using eq.~\ref{eq:P}, with a selection of the appropriate case for $p_\textrm{lensing}^c$ in eq.~\ref{eq:plensing}. $P_\textrm{star}$ is then just the sum of these individual probabilities. If we vary the mass $M$ of the PBHs, this changes the typical event duration. As a result, in addition to the overall probability $P_\textrm{star}$ changing as a function of $M$, the relative probabilities of these three event types will also change. We will utilize these probabilities for specific event types in sec.~\ref{sec:obs_dist}, when we try to compute the expected distribution of event observables for different types of lensing events.

\subsection{Modeling of the signal in the case where both the source and the lens are moving}
\label{subsec:lensmotion}

Our model of the AML signals in this section was based upon the assumption of a moving source and a static lens. Given that the typical random DM velocity in the halo is of a similar order as that of the typical stellar velocities, and given that the source lies in the background and that the PBH DM lies in the foreground, we would expect, on average, larger proper motion for the lens than for the source, which is counter to our modeling assumption in this section. One may therefore wonder how considering a motion of the lens will alter our calculations of the event duration and the lensing probability.

The displacement of the source image relative to the true position of the source depends only upon the relative angular separation of the source and the lens. This is an essential input to calculating the image position, which is valid both in the case where only the source is moving, as well as in the more general case where both the source and the lens are moving.

Let us suppose that the source is moving with a velocity $\vec{v}_s$ and the lens is moving with a velocity $\vec{v}_l$, relative to us. We will assume for simplicity that these velocities are perpendicular to our line-of-sight. Any velocity component parallel to our line-of-sight will not play any role in our lensing calculations. The relative angular separation of the source and the lens changes at a rate defined by the \textit{relative proper motion} between the two. The relative proper motion is given by,
\begin{eqnarray}
\vec{\mu} =  \frac{\vec{v}_s}{D_s} - \frac{\vec{v}_l}{D_l}.
\end{eqnarray}
The relative proper motion projected on to the lens plane thus yields an effective separation velocity parameter defined as,
\begin{equation}
\vec{v}_\textrm{sep} \equiv  D_l\times \vec{\mu} =  \vec{v}_s \frac{D_l}{D_s} - \vec{v}_l.
\end{equation}

Our calculation of the event duration and the lensing probability can be thought of as being computed in the frame of a hypothetical observer centered at the earth (or really at the sun, since we are ignoring parallax). However, if we think of this observer as rotating such that the lens maintains a fixed position in their field-of-view, then such an observer would see the kind of signal that we have assumed previously in this section, corresponding to a static lens  with $v_l = 0$. Although this rotation can never be realized by a true observer since we do not know the position of the PBH lens a priori, the event duration\footnote{Recall that the event duration is defined as the period of time for which the source and the lens separation is such that it gives rise to a displacement in the source position which is above \Gaia{} sensitivity threshold.} will be the same for the fictitious rotating observer and a true non-rotating observer, provided that we replace the parameter $v$ by $v_\textrm{sep} \equiv |\vec{v}_\textrm{sep}|$. Besides the event duration, the lensing probability calculations also remain unchanged provided we make this replacement of $v$ by $v_\textrm{sep}$.

Thus, the speed $v = 200$ km/s should be thought of as representing $v_\textrm{sep}$. Realistically, this replacement can not be fully correct because $v_\textrm{sep}$ depends upon the ratio $D_l/D_s$ which varies for different source-lens system. However, in lieu of a full modeling of the source and lens velocities, our modeling should be thought of as using a ``typical value'' $v_\textrm{sep} = 200$~km/s.

While the replacement of $v$ by $v_\textrm{sep}$ allows us to compute the event duration and the probability of lensing in the more general case of a moving lens, the key difference will be the actual signals seen by an observer who is not co-rotating with the lens as compared to the signals predicted in our modeling (see fig.~\ref{fig:EventDuration}). We have shown for comparision three different cases of source and lens velocities relative to us and the effect on the actual signal that would be seen by \Gaia{}.

\begin{figure}[h]
	\begin{subfigure}{0.31\textwidth}
		\includegraphics[width=\linewidth]{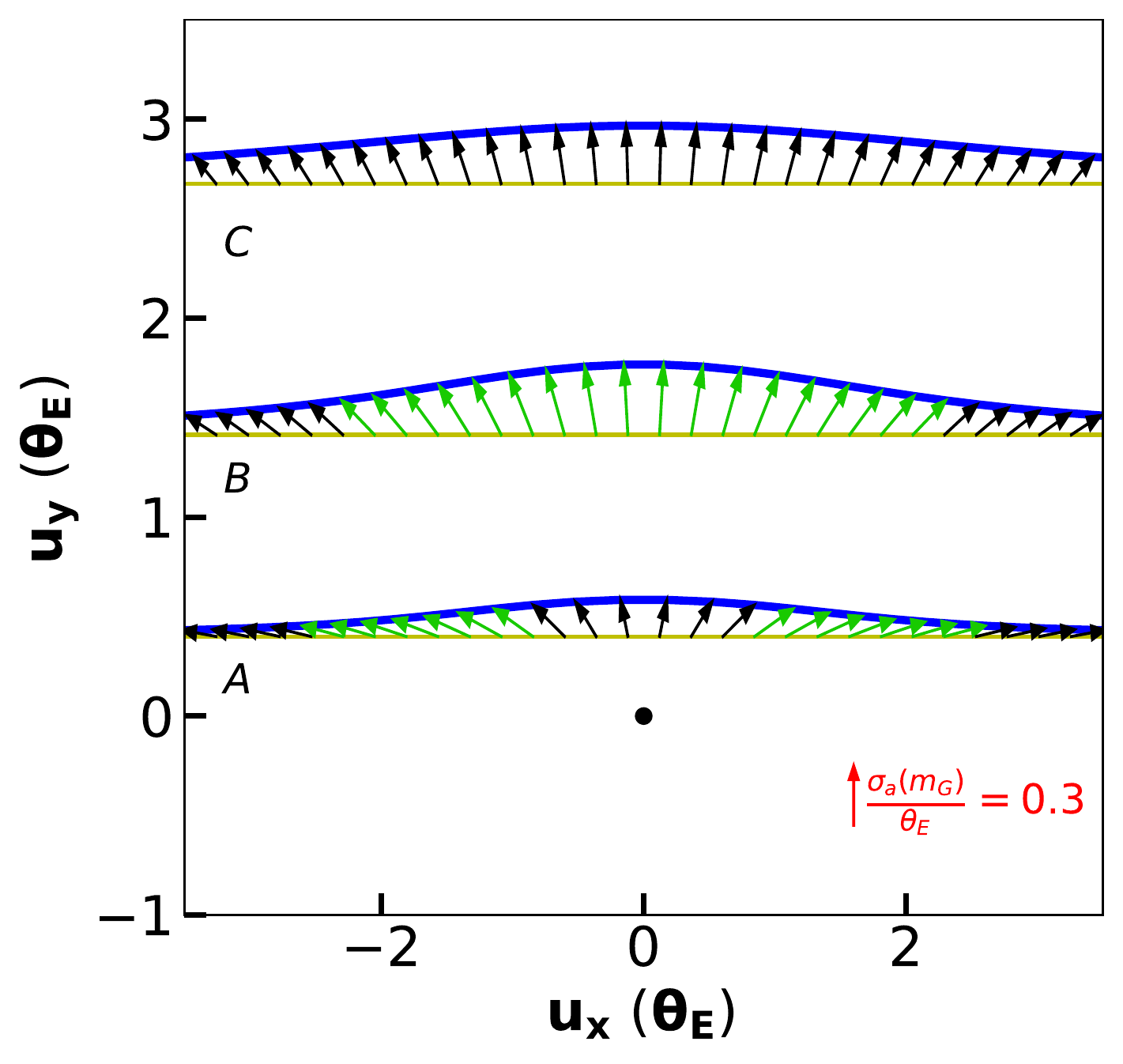}
		\caption{} \label{fig:Smoving}
	\end{subfigure}%
	\hspace*{\fill}   % maximize separation between the subfigures
	\begin{subfigure}{0.31\textwidth}
		\includegraphics[width=\linewidth]{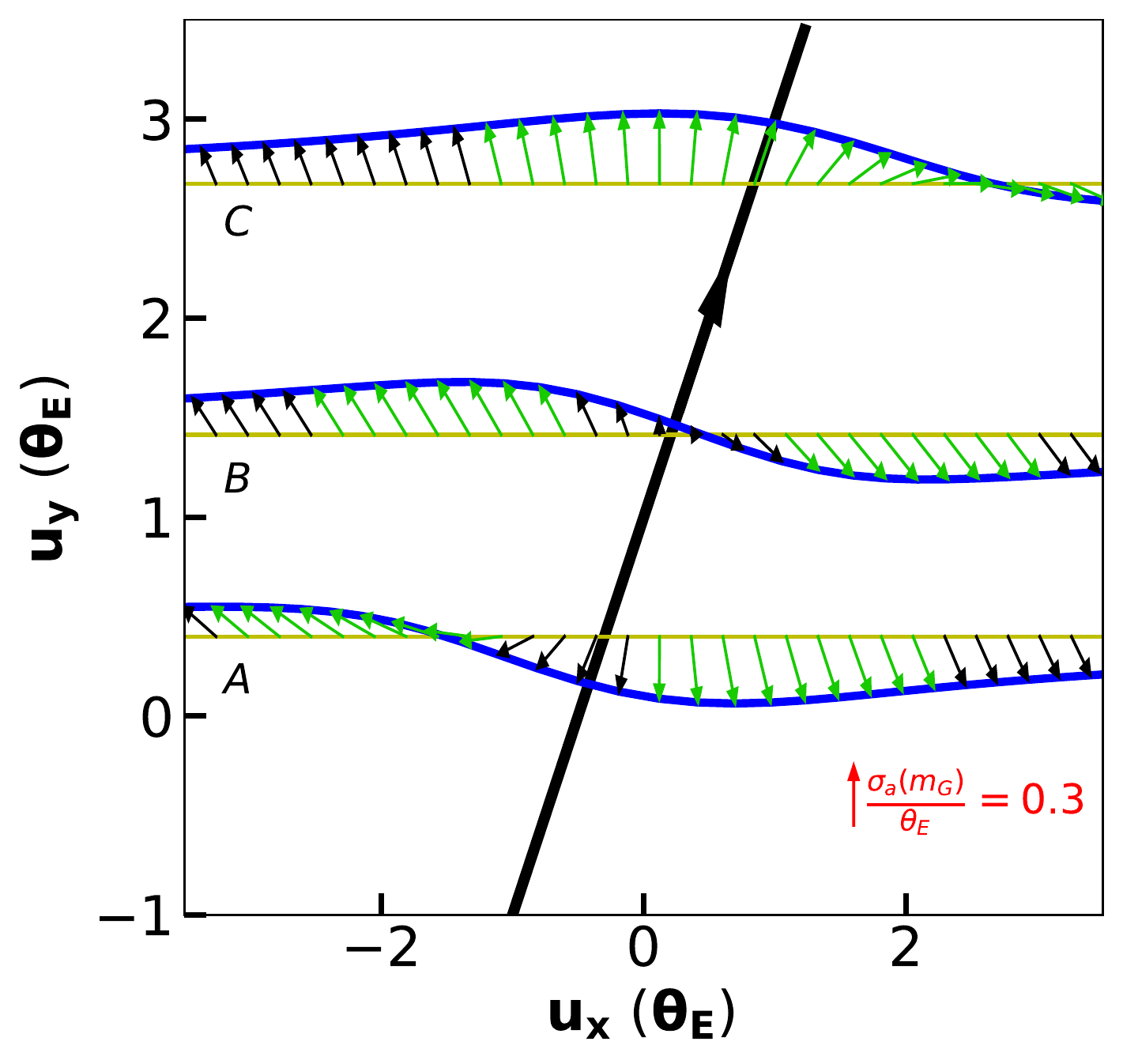}
		\caption{} \label{fig:LSmoving}
	\end{subfigure}%
	\hspace*{\fill}   % maximizeseparation between the subfigures
	\begin{subfigure}{0.31\textwidth}
		\includegraphics[width=\linewidth]{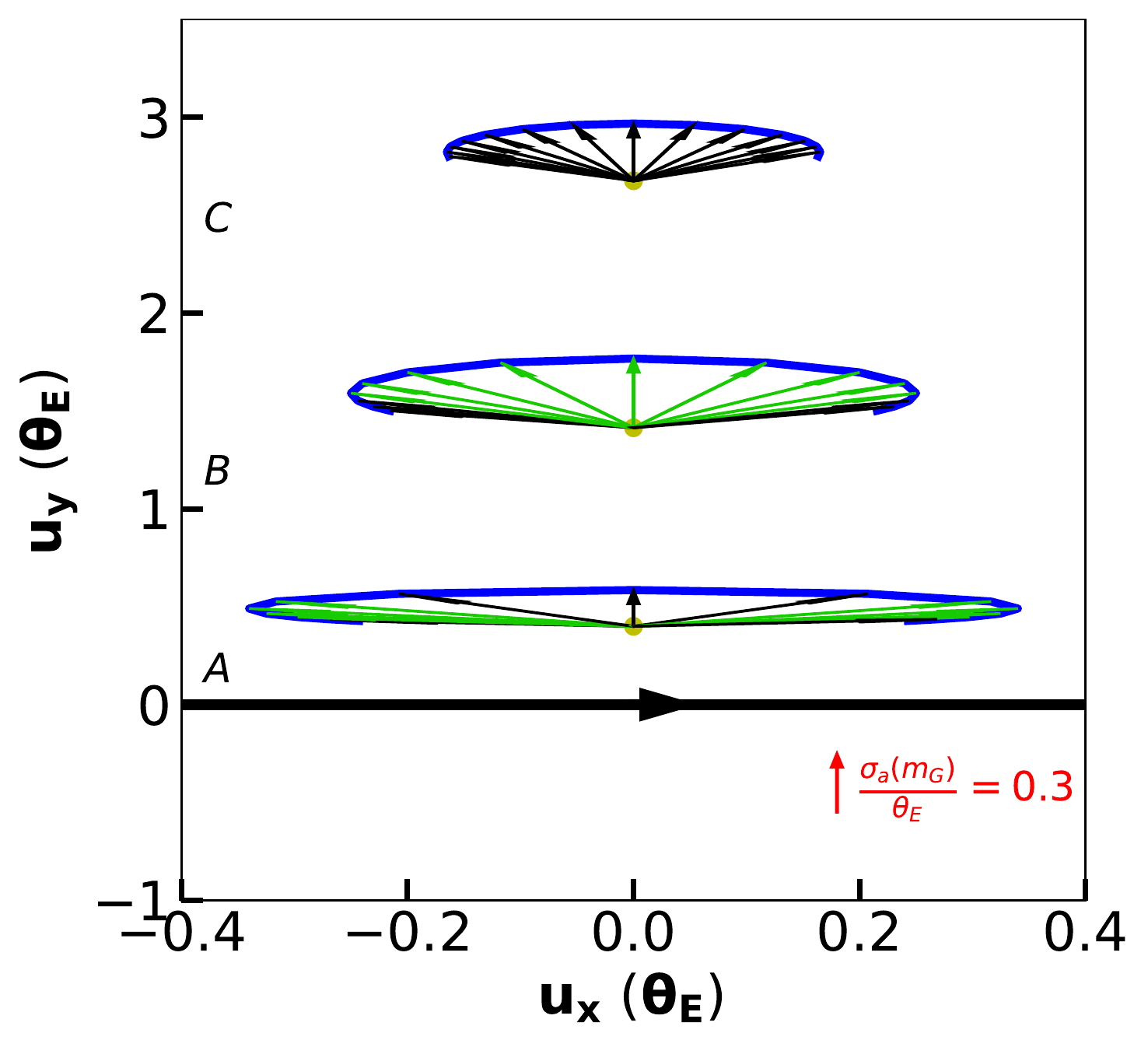}
		\caption{} \label{fig:Lmoving}
	\end{subfigure}
	\caption{This figure shows the realistic AML signals that would be seen for different cases of source and lens relative motion. In (a) the lens is static and the source is moving, in (b) both the source and the lens are moving, with the lens shown as the black solid line running from the bottom to the top of the figure, and in (c) the source is static and the lens is moving, with the lens trajectory indicated by the black horizontal line. In all three cases, the yellow line/yellow dot indicates the true source position, and the blue curve indicates the apparent position of the image due to the lensing effect. The black and green arrows indicate the displacement of the source image from its true position, with green arrows indicating displacements above threshold, and black arrows displacements below threshold. We can see from cases (b) and (c) in the figure that the actual signals seen by \Gaia{} can differ from our default modeling of moving source and static lens (case a).}
	\label{fig:RealSigvsModel}
\end{figure}

In fig.~\ref{fig:Smoving}, we show the case of a static lens (denoted by a black dot) with a moving source (whose true position is shown in yellow). The image of the source is shown in blue, with vectors indicating the displacement between the true positions and the corresponding image positions. As before we have shown the arrows in green when the astrometric shift is above \Gaia{} sensitivity threshold and in black when the astrometric shift is below threshold. This figure is identical to fig.~\ref{fig:EventDuration}.

In fig.~\ref{fig:LSmoving}, we show exactly the same quantities (true positions of the source in yellow, apparent positions of the image in blue, and displacement vectors in black and green), however in this case the lens has a rectilinear motion of its own and its trajectory is shown by the black line which extends from the bottom to the top of the figure.

In fig.~\ref{fig:Lmoving}, we show the case of a static source with a moving lens (whose trajectory is depicted as the horizontal black line). In this case, the true source position is static but the image traces a closed loop as the lens passes in the vicinity of the source~\cite{1995ApJ...453...37W}.

For all three cases, we have shown three different source trajectories, each of which corresponds to a different choice of impact parameter. Depending upon the event duration, for each case, a part of, or the entirety of, the trajectory may be seen by \Gaia{}. As we have argued above, each of these cases will give rise to the same event duration if the parameter $v_\textrm{sep}$ is the same. Moreover, the probability of detecting a lensing signal in each of these cases is also the same given a fixed parameter $v_\textrm{sep}$. However, the actual signal that \Gaia{} will see is clearly very different in each of these three cases. Thus, while the choice of our modeling of selecting a static lens and a moving source does not affect our prediction of signal event rate at \Gaia{}, it does not adequately describe the types of signals that will be seen by the telescope. The reason why this issue is important is because in order to predict the background rate and how well the background can be separated from the signal, we need to know what the actual signals look like in the data.

Given the fact that the source lies behind the lens, one might typically expect that the lens proper motion dominates over source proper motion in contributing to the relative proper motion. If this were true then we would expect realistic signals that look closer to those of fig.~\ref{fig:Lmoving} rather than those of fig.~\ref{fig:Smoving}. However, we caution that a full numerical study which includes a probabilisitic distribution of source and lens velocities is necessary to determine what the typical values of $v_\textrm{sep}$ are and what the relative importance of source proper motion versus lens proper motion are in the expected signal sample that \Gaia{} will be able to detect. Only then one can make definitive claims about which of the cases in fig.~\ref{fig:RealSigvsModel} is more representative of the typical signals that \Gaia{} will be able to see.

In sec.~\ref{sec:background}, where we discuss various sources of background, we will make the assumption that the lensing signals are of the type in fig.~\ref{fig:Smoving}. Since our analysis of the background rate will be highly qualitative, we do not expect this assumption to have a significant effect on our estimate of this rate, however a more quantitative study of the background rate would actually have to take in to account the relative fraction of the different types of lensing signals. This study would perhaps best be performed numerically.

\subsection{Including the effect of parallax}
\label{subsec:parallax}
 The analysis of this section has one major omission, which is the neglect of parallax. When looking for a lensing signal, we would have to look for deviations of the source motion from not just rectilinear motion but from rectilinear plus parallactic motion. The parallax of the source itself is an unknown parameter that needs to be fitted for. One would have to define a parameter (such as the best fit $\chi^2$) which captures the difference between the observed trajectory and an expected trajectory with rectilinear plus parallactic motion only. A large value of this parameter would indicate a significant deviation, which could perhaps be attributed to a lensing signal.

 The types of lensing signals that would be seen by \Gaia{} would also be different from the expectations of fig.~\ref{fig:RealSigvsModel} because of the omission of parallax. Since the lens is closer to us than the source, we would expect the lens to have a larger parallax than the source. Lens parallax would significantly complicate the types of deviations seen in fig.~\ref{fig:RealSigvsModel}.

 A numerical study which includes the effect of parallax is needed to see the effect on the event duration and the probability of lensing. Moreover, once we understand the full trajectories including parallax, both with and without lensing, we can then try to see which backgrounds can mimic lensing signals, and to see if there are new discriminatory variables that can separate genuine lensing signals from these backgrounds. We leave such a study to future work.

\section{Lensing signal rate due to PBHs}
\label{sec:lensingrate}
Now we are ready to numerically compute the expected total number of lensing ($N_l$) events that \Gaia{} will see over a time period $t_\textrm{obs}$, for a given hypothesis of the PBH parameters $(f,M)$.  We have seen in sec.~\ref{sec:LensingProb} that there are three kinds of AML signals that can be detected by \Gaia{}. However, the SDLE event rate is typically highly suppressed by the coincidence probability for most PBH masses, and more importantly, as we will show in sec.~\ref{sec:background_stat} SDLEs will be completely dominated by the background due to centroiding uncertainty. Thus, in this section we will focus only on the case of IDLEs and LDLEs. The numerical results for the computation of $N_l$ will be shown in sec.~\ref{sec:lensingrate_numerical}. Later, in sec.~\ref{sec:obs_dist}, we will also introduce some simplified event observables for IDLE and LDLE type events, and compute the expected distribution of these observables.

\subsection{Computing the number of lensing events that \Gaia{} will see}
\label{sec:lensingrate_numerical}

\begin{figure}
    \centering
    \includegraphics[scale=0.45]{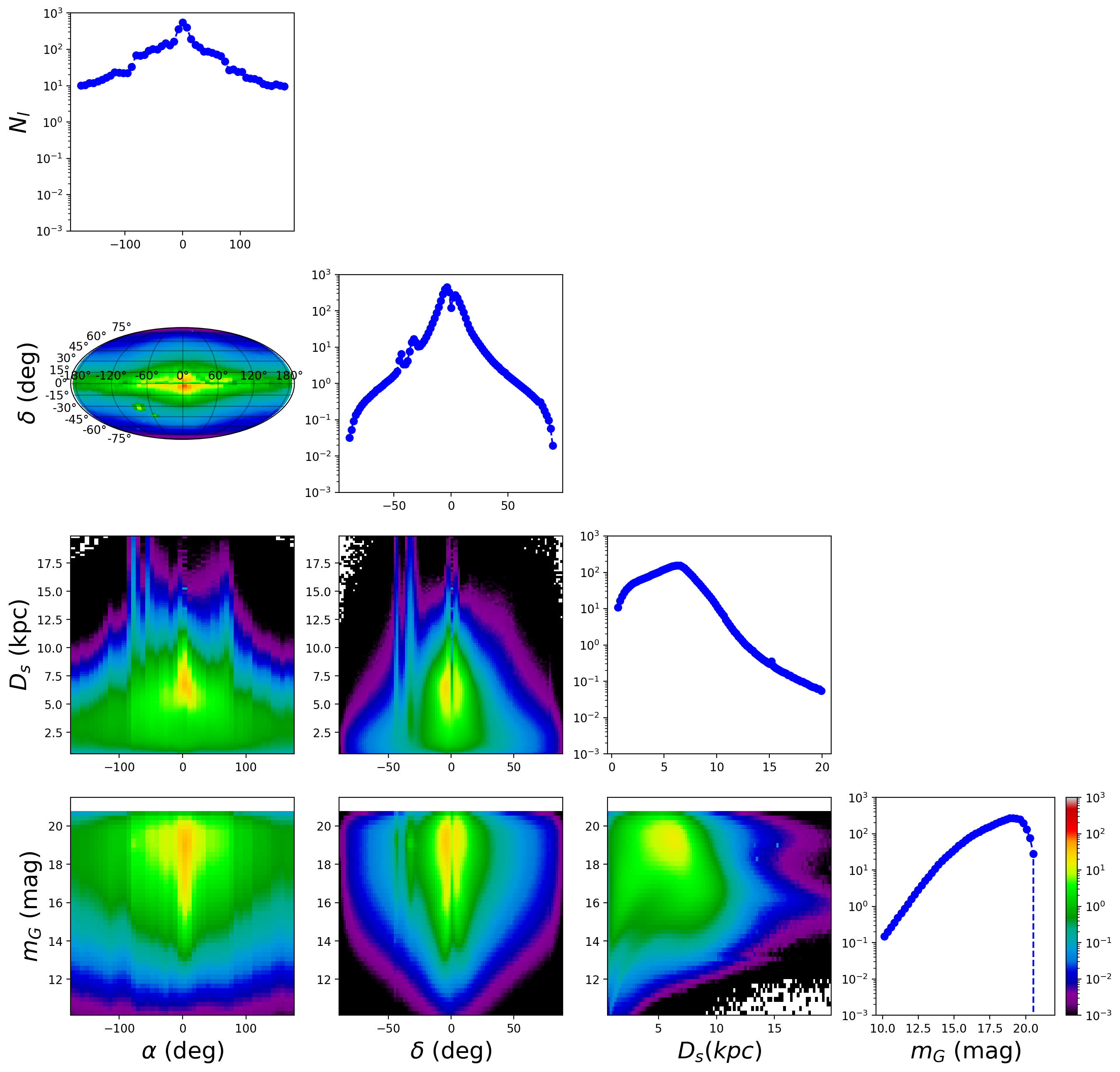}
    \caption{Differential distribution of the expected number of lensing events ($N_l$) for PBH parameter $(f,M) = (1,~14~M_\odot)$. The total number of detectable lensing events is $\sim 3725$, with 55\% of these being LDLE type, and the rest IDLE type (we have once again neglected SDLE type events which will have a highly suppressed rate due to the coincidence probability). $N_l$ is obtained by weighting the distribution of stars in the \Gaia{} catalog by the probability $P_\textrm{star}$ (eq.~\ref{eq:P}) for a star to undergo a detectable lensing event. $N_l$ depends on the location of stellar targets for AML searches ($\alpha,\delta,D_s$) and the apparent magnitude $m_G$ of these sources. The figure shows 2D as well as 1D distributions of $N_l$ as a function of these parameters. From the figure we can see that the maximum number of lensing events will be obtained for stars in the direction of the Galactic center ($\alpha = \delta = 0^\circ$). The most interesting part of this figure is the 1D dependence of $N_l$ on the apparent brightness $m_G$ of target stars (bottom right panel). This dependence arises due to an increase in the number of stars at low apparent brightness (high $m_G$) weighted by a higher probability of a detectable lensing signals from stars with higher apparent brightness (low $m_G$).}
    \label{fig:NlGeDR3Coarse-Grained}
\end{figure}

We start by fixing a hypothesis for the PBH parameters $(f,M)$.
In order to calculate the expected total number of lensing events  that \Gaia{} will see over $t_\textrm{obs}= 5$~years, we need to compute the average number of lensing events that we expect to see for each \emph{individual} star in the \Gaia{} catalog, and then add up all these values.

For each star in our reduced \Gaia{} catalog, described in sec.~\ref{subsec:GaiaEDR3Catalog}, we have $D_s$, $\alpha$, $\delta$, and $m_G$ values. We can then numerically calculate the probability $P_\textrm{star}$ (eq.~\ref{eq:P}) that a lensing event will occur and will be detected by \Gaia{} in $t_\textrm{obs} = 5$ years. This probability can also be interpreted as the average number of lensing events associated with a given star. Since, the computation for $P_\textrm{star}$ was numerically challenging, we take the nearest grid point in $D_s$, $\alpha$, $\delta$, and $m_G$ that we have tabulated $P_\textrm{star}$ values for, to determine its value. We then sum over all the stars in the \Gaia{} catalog to determine the average expected number of detectable lensing events. This is equivalent to a weighting of $P_\textrm{star}$ by the distribution of stellar locations and apparent magnitudes in the Galaxy, or equivalently, for the choice of $f = 1$ and $M = 14~M_\odot$, it is a weighting of fig.~\ref{fig:LensingEvents} by  fig.~\ref{fig:GeDR3grid50}. The result of this procedure is shown in fig.~\ref{fig:NlGeDR3Coarse-Grained}, in which we plot the total number of detectable lensing events as a function of $D_s$, $\alpha$, $\delta$, and $m_G$.

From fig.~\ref{fig:NlGeDR3Coarse-Grained}, we can see that the location on the sky where the maximal number of lensing events will occur is in the direction towards the Galactic center. This is because of a combination of a large number of stars along with a large DM density integrated along the line-of-sight in this direction. A more interesting trend is the dependence of the number of lensing events on $m_G$. As $m_G$ increases, \Gaia{}'s astrometric positioning errors also increase (see eq.~\ref{eq:Asigma}), and hence the lensing probability $P_\textrm{star}$ decreases. However, there are a larger number of stars with higher values of $m_G$ in the \Gaia{} catalog (see fig.~\ref{fig:GeDR3grid50}). The competition between these two opposing effects is what leads to the non-trivial dependence on $m_G$, leading to a peak in the distribution of number of lensing events near $m_G \simeq 19$~mag.

\begin{figure}
    \centering
    \includegraphics[scale=0.7]{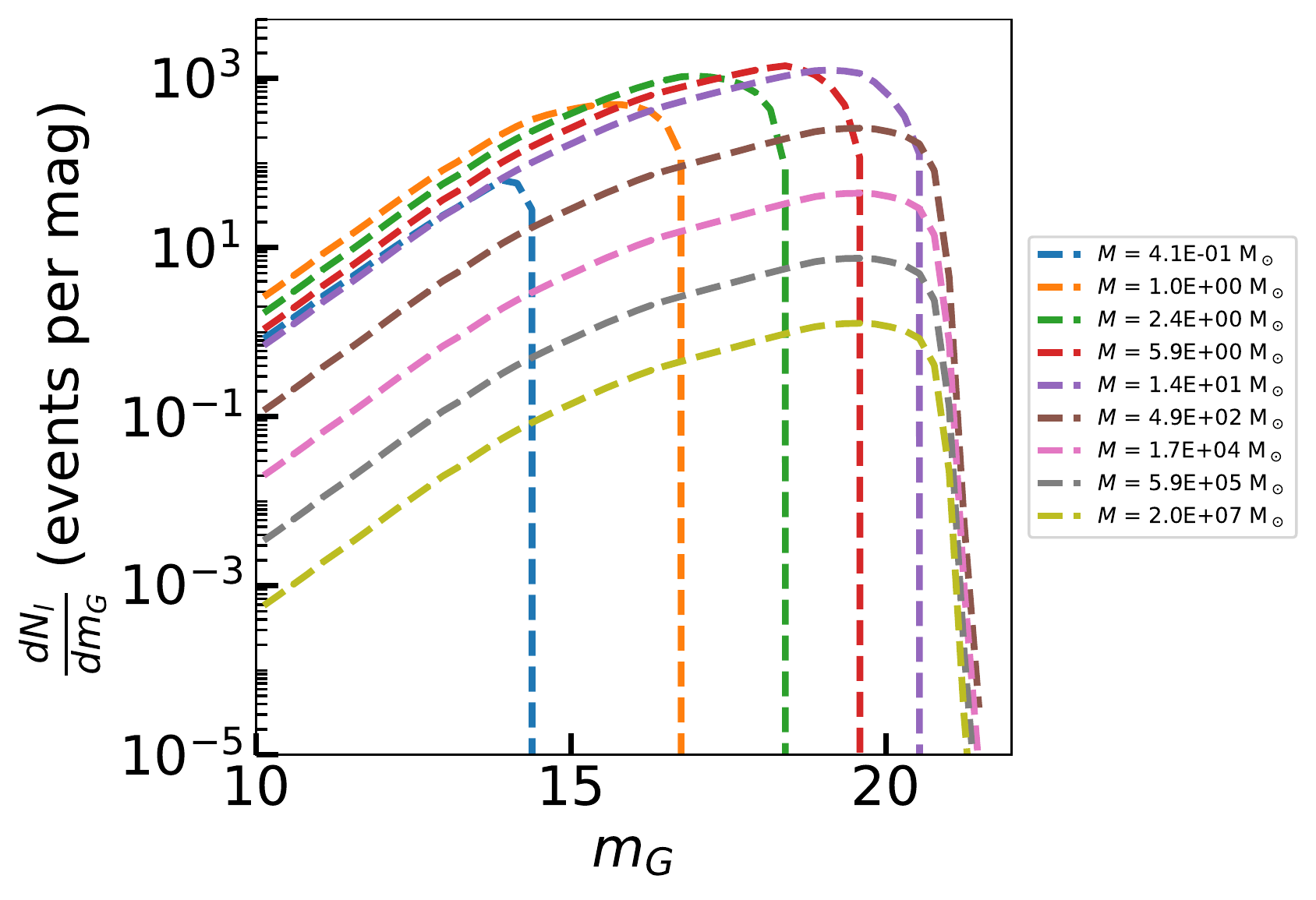}
    \caption{ 1D distribution of the expected number of lensing events for stars in the \Gaia{} catalog as a function of the target star's apparent brightness ($m_G$). We have shown this distribution for several values of the assumed PBH mass $M$, while assuming that they make up a fraction $f=1$ of the Galactic DM. For PBH masses lower than $10^{-2}~M_\odot$, the number of expected events drops rapidly to zero as both the deflection and event duration become small. For larger PBH masses the number of potential lenses decreases, which leads to fewer lensing events with increasing PBH mass.}
    \label{fig:mGDist}
\end{figure}

We can also try to understand how the distribution of event numbers will change as a function of the PBH mass $M$. Since the number of events is still expected to peak in the direction of the Galactic center, we will focus our attention on the $m_G$ dependence of the expected number of events. In fig.~\ref{fig:mGDist}, we show for different PBH masses $M$, the 1D dependence of the expected number of events on $m_G$, after marginalizing over all $D_s$, $\alpha$, and $\delta$ values. At low values of $M \lesssim~10^{-2} M_\odot$, the expected number of events drops rapidly to zero. This is because of a combination of several factors: (i) the typical AML event duration drops below \Gaia{}'s sampling time $t_s$ and (ii) the typical AML deflection drops below \Gaia{}'s sensitivity threshold $\sigma_a(m_G)$. The combination of these two factors makes the AML signals unobservable at \Gaia{} for low PBH masses. For large masses the expected number of events drops more smoothly, as can be seen in the figure. This drop in the number of lensing events at large masses is because there are fewer lenses for a fixed PBH mass density, and hence the probability of AML lensing decreases.

\subsection{Expected distribution of observables for AML events}
\label{sec:obs_dist}

In the case of IDLEs one would observe a rectilinear trajectory with a distortion or deviation from apparent rectilinear motion for a short time, i.e. for the duration of the event $t_e$. Two simplified observables can characterize such trajectories: the event duration $t_e$ during which the deflection from rectilinear motion is above \Gaia{}'s threshold, and the maximum deflection $(\delta \theta)_\textrm{max}$ of the trajectory away from rectilinear motion. For LDLE type events it is harder to find a simplified observable that would truly characterize such events. However, we can try to examine the relative deflection difference between the start point and the end point of the trajectory away from the true rectilinear trajectory. Note that for LDLE type events, the event duration $t_e$ will not be an easily reconstructable observable. Given our computation of $P_\textrm{star}$ in the previous section, and more specifically, the computation of $P_\textrm{star}^\textrm{IDLE}$ and $P_\textrm{star}^\textrm{LDLE}$, we can plot the expected distribution of these simplified observables, for a given hypothesis of PBH parameters.

\subsubsection{IDLE observables}

First, we need to compute, for a star located at $D_s, \alpha, \delta$, and with apparent brightness $m_G$, the average expected values of the  observables $t_e$ and $(\delta \theta)_\textrm{max}$.

We had earlier evaluated the average event duration $\langle t_e \rangle$ over impact parameters in eq.~\ref{eq:eventduration}. We further average this over all candidate IDLE lensing events that such a star will undergo, due to lenses at various distances $D_l$ between us and the star along the line-of-sight. This gives us,
\begin{align}
\label{eq:te_avg_Dl}
  \langle\langle t_e \rangle\rangle & = \frac{\int_0^{D_s} \langle t_e \rangle \frac{dP_\textrm{star}^\textrm{IDLE}}{d D_l}dD_l }{\int_0^{D_s} \frac{dP_\textrm{star}^\textrm{IDLE}}{d D_l}dD_l},
\end{align}
where the double angular brackets around $t_e$ denote averaging with respect to both -- impact parameters as well as candidate lens distances.

We then compute a similar double average for $(\delta \theta)_\textrm{max}$. The averaging over impact parameters can be performed analytically, and is given by,
\begin{align}
    \langle  (\delta \theta)_\textrm{max} \rangle & = \theta_E\frac{1 + \log \left[\frac{1}{8}K(K+\sqrt{K^2 - 8})\right]}{K+\sqrt{K^2 - 8}},
\end{align}
where $K=\frac{\theta_E}{\sigma_a(m_G)}$. The derivation of this equation is given in appendix~\ref{sec:appendixObs}. We can then perform a similar further averaging as we did for $t_e$, for this observable, over lens distances $D_l$.

\begin{figure}[ht]
\centering
\hspace*{-0.95in}
\begin{subfigure}{.4\textwidth}
  \centering
  \includegraphics[scale=0.45]{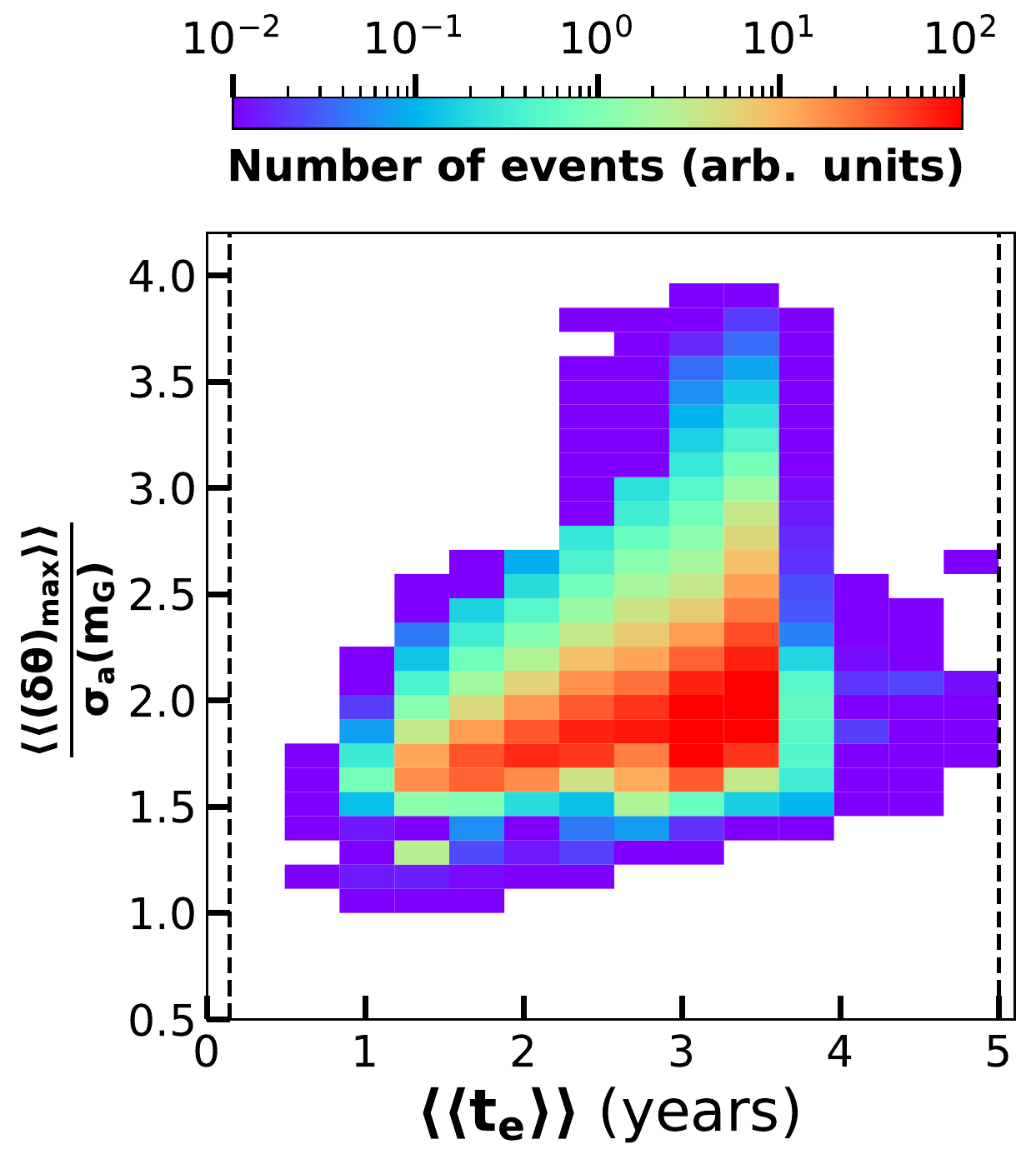}
  \caption{IDLE observables}
  \label{fig:sub-first}
\end{subfigure}
\begin{subfigure}{.4\textwidth}
  \centering
  \hspace*{0.05in}
  % include second image
  %\includegraphics[scale=0.5]{plots/DsGeo/rho0modifiedAvgte_DoubleAvgdeltamaxDavgPstarLDLEGeDR3Empirical.png}
  \includegraphics[scale=0.45]{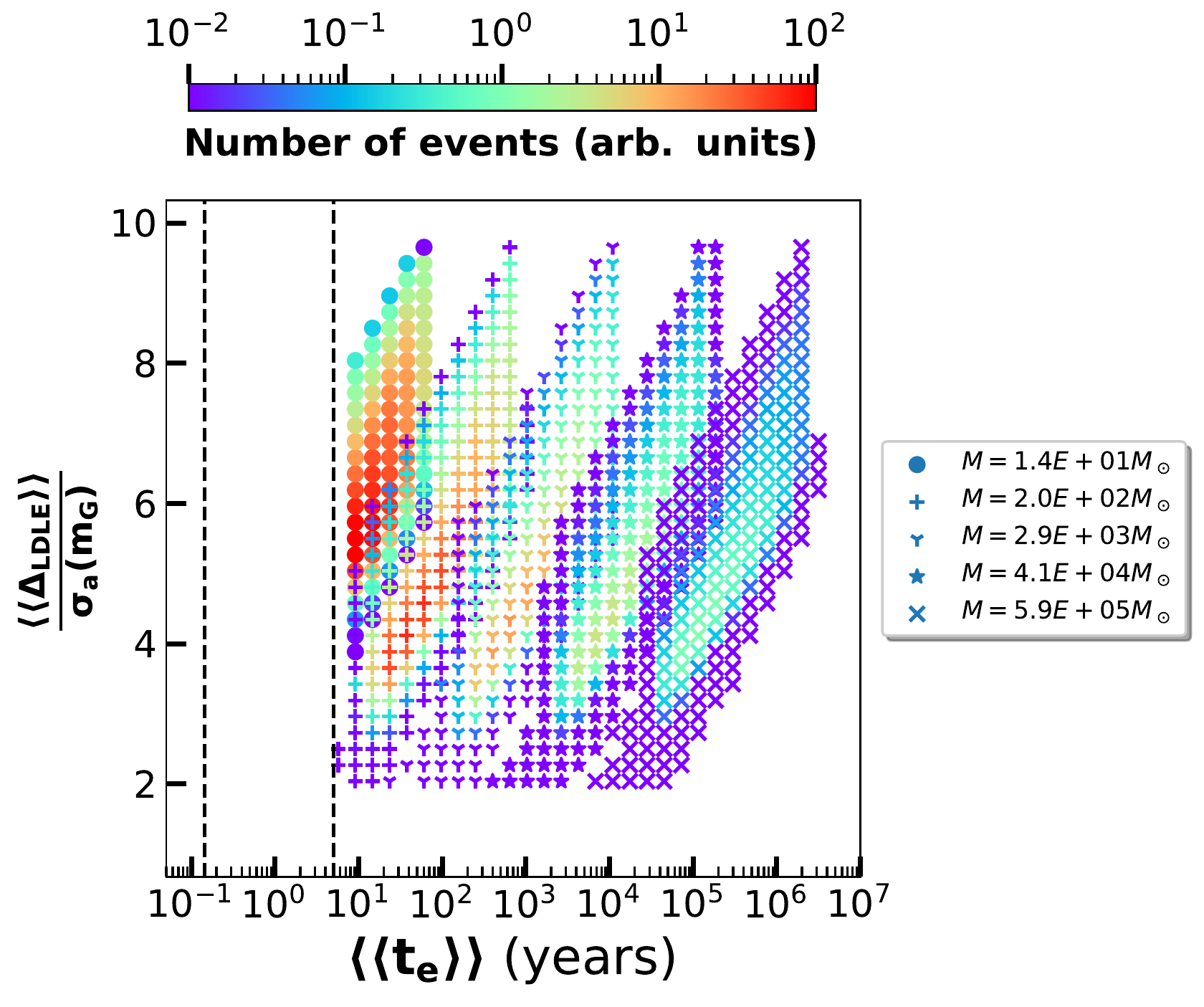}
  \caption{LDLE observables}
  \label{fig:sub-second}
\end{subfigure}
\caption{The two histograms above show the simplified distribution of IDLE and LDLE observables. For IDLEs, we have plotted the distribution for a PBH mass of $14~M_\odot$, whereas we have shown the LDLE observable distribution for several different PBH masses. IDLE event durations lie between \Gaia{}'s sampling time $t_s$ and observational time $t_\textrm{obs}$. LDLE event durations are larger than $t_\textrm{obs}$ and are thus not easy to extract from an observed trajectory.}
\label{fig:ObsDist}
\end{figure}

We now compute these double average observables for each star in the \Gaia{} catalog and construct a 2-D histogram of $\langle\langle t_e \rangle\rangle$ and $\langle\langle (\delta \theta)_\textrm{max} \rangle\rangle$. While populating the histogram, we weight each star's contribution by $P_\textrm{star}^\textrm{IDLE}$, i.e. the probability that such a star will undergo an IDLE type event. The resulting histogram shows the expected distribution of the observables for IDLE type events. This histogram is shown in fig.~\ref{fig:sub-first} for $M=14~M_\odot$. We have also shown with vertical dashed lines the sampling time $t_s = 52.2$~days and the observational time $t_\textrm{obs}=5 $~years.

From the figure we can see that the distribution of typical IDLE type event durations are between 1.5~-~4~years at this mass. These event durations lie between $t_s$ and $t_\textrm{obs}$. However, for lower choices of mass for the PBH, the distribution of event durations will drift to lower values, and eventually for sufficiently small masses, $M\lesssim 10^{-2}~M_\odot$, would fall below the sampling time for \Gaia{}. Thus, such low mass PBHs would typically not induce an IDLE lensing signal (although they might induce SDLE type events). As we shall see in the next section, the statistical backgrounds are significant for event durations $t_e \lesssim 2$~years. We will demand that the IDLE signals we are looking for have event durations greater than 2 years so that this background is negligible, and this will result in \Gaia{} only being sensitive to PBHs with masses $M\gtrsim 0.1~M_\odot$.

\subsubsection{LDLE observables}
In the case of LDLE type events, the relative deflection difference observable, averaged over impact parameters, is computed in appendix~\ref{sec:appendixObs}, and is given by,
\begin{equation}
    \langle \Delta_\textrm{LDLE} \rangle = \frac{v t_\textrm{obs} }{D_l} \frac{2\textrm{log}(u_T/u_\textrm{min})}{u^2_T - u^2_\textrm{min}},
\end{equation}
where $u_T = \sqrt{ \frac{ v t_{\text{obs} }  }{D_l \sigma_a(m_G)} } $ and $u_\textrm{min}=\sqrt{2}$. We can then further average this over $D_l$ values, using a formula similar to \ref{eq:te_avg_Dl}, however this time with $P_\textrm{star}^\textrm{IDLE}$ replaced by $P_\textrm{star}^\textrm{LDLE}$.

We could now build a 1D histogram for this observable by populating it with all the stars in the \Gaia{} catalog, appropriately weighting each star's contribution by $P_\textrm{star}^\textrm{LDLE}$, i.e. the probability for that star to undergo an LDLE. However, instead of simply computing the 1-D distribution of $\Delta_\textrm{LDLE}$, we compute the 2-D distribution of  $\Delta_\textrm{LDLE}$ \textit{and} $t_e$. Although $t_e$ will not be a genuine direct observable for LDLEs, it can still be thought of as a parameter characterizing the trajectory, albeit one that is much harder to extract. Furthermore, it has a straightforward comparison to $t_e$ for IDLE type events. The 2-D histogram for these observables for LDLE type events is shown in fig.~\ref{fig:sub-second} for several values of the PBH mass $M$ ranging from $14~M_\odot$ to $5.9 \times 10^5~M_\odot$.

Higher values of the PBH mass lead to larger event durations, but they give a similar distribution of the deflection observable. However, the total number of events decreases at large PBH masses because of the decrease in number of PBHs in the Galaxy for a fixed DM density. This decrease will limit the sensitivity of \Gaia{} to very high mass PBHs.

Note that for a given PBH mass, such as for $14~M_\odot$ as shown in the figure, it is possible to have both IDLE and LDLE type events. Ideally the distribution of $t_e$ should look continuous as we transition from IDLE to LDLE type events in figs.~\ref{fig:sub-first} and~\ref{fig:sub-second}, but the lack of smooth continuity is because of our assumptions of very large event durations $t_e \gg t_\textrm{obs}$, when deriving the probabilities and event rates for LDLE type events.

\section{Background rate}
\label{sec:background}

In this section we will discuss various sources of background that can mimic the AML signal caused due to PBHs. These backgrounds can lower the statistical significance of a potential PBH signal, and therefore careful modelling is required to estimate the background rate. We can categorize the backgrounds into three types: statistical, systematic, and astrophysical. The statistical background is due to uncertainty in the centroiding of the image of a given star. The systematic backgrounds are due to instrumental systematics. The astrophysical backgrounds are due to i) astrophysical lenses which can mimic the effects of PBH lensing, and ii) deviation from our assumption of rectilinear motion, for example due to a binary companion. A detailed study of all the backgrounds is beyond the scope of this work. We will however discuss the statistical and astrophysical backgrounds. For the statistical backgrounds we will estimate their rate for the case of IDLEs and compare it to our predicted event rates from PBHs. We will use this estimate to argue that the background rate for SDLEs is so large that it makes SDLE event detection at \Gaia{} nearly impossible. We will not estimate the astrophysical background rates, but we will discuss which sources of this background we expect to be relevant, and some possible ways to reduce these.

In the previous section we showed that the lensing signals can be characterized by a few observables. For example, for IDLEs, we would have the event duration $t_\textrm{e}$ during which the deflection is above \Gaia{}'s threshold sensitivity, and the maximum astrometric shift $(\delta \theta)_\textrm{max}$, whereas for LDLEs, we would have $\Delta_\textrm{LDLE}$. In addition, for both these types of events, we would also have the location in the sky $(D_s,\alpha, \delta)$ of the background star, where $D_s$ could be measured through parallax. Under a specific hypothesis of the PBH parameters $(f,M)$, we have computed the expected distribution of a few of these simplified observables for genuine AML signals from PBHs.
In principle, one could also augment these astrometric observables with \Gaia{}'s relatively crude photometric measurements to also characterize the PML magnification signal and predict the expected distribution. Each candidate lensing event can thus be thought of as a single point in the multi-dimensional space of these observables.

In general we expect that the  backgrounds that could mimic an AML signal will have a different distribution in this space as compared to that of genuine AML signals. Thus, by applying suitable cuts on an observed distribution in this space, we should be able to enhance the signal-to-background ratio.

\subsection{Statistical Backgrounds}
%\label{subsec:BckStat}
\label{sec:background_stat}
Let us first discuss the case of statistical background for IDLEs. For an unlensed trajectory, since the centroiding of a star in \Gaia{}'s detectors has an intrinsic uncertainty of $\sigma_a (m_G)$, this would lead to statistical fluctuations in the observed position of a star. If these statistical fluctuations mimic the lensing signal due to PBHs, then they would form a statistical background.

While estimating the lensing probability for a given star $P_\textrm{star}$ due to PBHs, we assumed that the instantaneous astrometric deflection of a star's trajectory $\delta \theta$ is greater than \Gaia{}'s threshold sensitivity $\sigma_a (m_G)$. If the event duration is $ t_e $, then there would be $N_s \simeq  t_e /t_s$ observations of the star's position with an above threshold lensing signal, where $t_s$ is \Gaia{}'s sampling time.

We can thus estimate the rate at which the  statistical fluctuations of an unlensed trajectory generate a fake IDLE type lensing signature using the critera above. For the statistical fluctuations to mimic a lensing signal with event duration $t_e$ (where $t_s < t_e < t_\textrm{obs}$), we need $N_s$ consecutive fluctuations in the centroiding, each of which are greater than $1$-$\sigma$ away from the true rectilinear trajectory. Here $N_s = t_e/t_s$ is the number of passes that \Gaia{} would make of the star during a time $t_e$. Moreover, these fluctuations all need to be in the same direction to mimic a lensing signal. If we assume that, for a given star in the \Gaia{} catalog, the centroiding fluctuations in a single pass are governed by a gaussian distribution with standard deviation $\sigma_a (m_G)$, then the probability of greater than $1$-$\sigma$ single-sided deviation is $1-\textrm{erf}(1) \simeq 0.16$, where $\textrm{erf}$ is the error-function. Thus, for a typical trajectory with $35$ samples of the star's position as measured by \Gaia{} over $t_\textrm{obs}=5$~years, we would find that the probability for a star to undergo such a fake event due to statistical fluctuations is,
\begin{equation}
	\label{eq:Pstat}
    P_\textrm{star}^\textrm{stat} = 2\times(0.16)^{N_s}(35-N_s),
\end{equation}
where $0.16^{N_s}$ is the probability for $N_s$ $1$-$\sigma$ single-sided deviations in the star's apparent trajectory, the factor of $2$ takes into account that the deviations can be on either side of the trajectory, and the factor of $35-N_s$ gives the number of ways of choosing the starting pass for which the fake lensing event begins. $P_\textrm{star}^\textrm{stat}$ does not depend on the specific properties of the star such as its location and apparent magnitude. Thus, we can directly multiply it by the number of stars in the \Gaia{} catalog to obtain the expected number of statistical background events.

\begin{figure}
	\centering
	\includegraphics[scale=0.7]{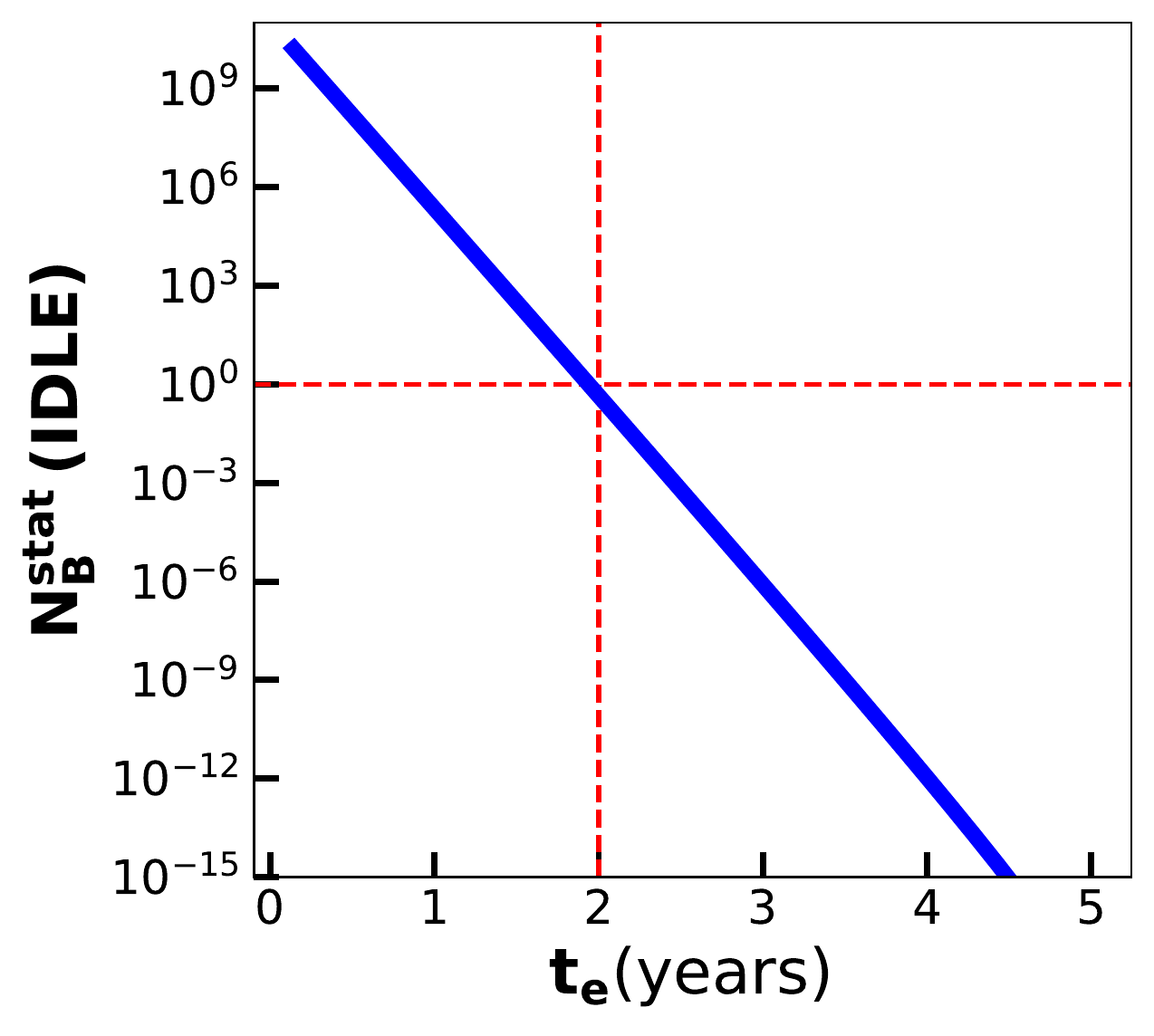}
	\caption{This plot shows the estimate of the number $N_\textrm{B}^\textrm{stat}$ of statistical background events for IDLEs and its dependence on the duration $t_e$ of AML events. Here, $N_\textrm{B}^\textrm{stat} = P_\textrm{star}^\textrm{stat} \times1.47\times10^9 $, where $P_\textrm{star}^\textrm{stat}$ is computed in eq.~\ref{eq:Pstat}. We have also shown guiding lines (dashed-red) to indicate an event duration of 2 years for which the statistical background is $\sim$ 1 event. For event durations larger than this, the statistical background for IDLEs is nearly negligible.}
	\label{fig:StatBackground}
\end{figure}

The fake rate is obviously very small for longer event durations ($t_s \ll t_e \lesssim t_\textrm{obs}$ or $1\ll N_s \lesssim 35$) because of the requirement that multiple consecutive passes show large deflections, for e.g. for $t_e \simeq$~2~years, we find $N_s = 14$, and  $P_\textrm{star}^\textrm{stat} = 3\times 10^{-10}$. Upon multiplying this by the number of stars in the \Gaia{} catalog ($1.47\times10^9$), we would obtain $\sim 0.4$ background events with a 2 year event duration.  For smaller event durations, we would find a much larger number of background events. We have plotted the number of statistical background events for IDLEs as a function of the event duration $t_e$ in fig.~\ref{fig:StatBackground}. In order to reduce the statistical background to negligible levels, we make the simple choice to only look at IDLEs with event duration greater than 2 years.

We can also now understand why the statistical backgrounds for SDLEs would be large from the figure. As we can see, the number of background events increases rapidly for small event durations. Thus, we expect to get a very large number of statistical background events for SDLEs as well, which could correspond to an astrometric fluctuation in the position of the source image in a single pass through the entire trajectory. This validates the claim that we have made earlier that SDLEs will be undetectable at \Gaia{} due to the overwhelmingly large statistical background.

Here, we have looked at the dependence of the background rate on only one observable, the event duration $t_e$ and discussed the distribution of this observable due to statistical background. It is conceivable that if we examine the multi-dimensional space of observables, the background could be reduced further, thus allowing for relatively low background searches for IDLE signals with event durations slightly shorter than 2 years.

For LDLE type events, estimating the statistical background is trickier. The reason for this is that there is no reference rectilinear motion against which we can study a deviation. Instead, in a numerical study we could simulate several rectilinear trajectories with statistical fluctuations, and compute the probability of some of these trajectories to be statistical outliers, perhaps through quantifying their $\chi^2$ fit to a pure rectilinear trajectory. Such a detailed numerical study is beyond the scope of the present work, but we hope to take it up in a future work.

\subsection{Astrophysical background}
\label{sec:background_astro}
There are several astrophysical sources of background that can mimic the lensing signatures due to PBHs, i)~astrophysical lenses which can mimic the effects of PBH lensing, ii)~ deviation from our assumption of rectilinear motion, for example due to a binary companion. We will only briefly discuss these here, along with some suggestions for background reduction techniques.

Besides AML events due to PBHs, \Gaia{} is also expected to see AML events due to microlensing caused by foreground stars~(including objects formed from stellar collapse)~\cite{2002ApJ...576L.131A,2002MNRAS.331..649B, 2018A&A...618A..44B}, brown dwarfs~\cite{Belokurov:2001vh, 2018A&A...620A.175K, 2018AcA....68..351N}, and free-floating planets~\cite{2018Ap&SS.363..153H}. Such events form a background to the detection of PBHs through their AML signatures. These objects would predominantly lie in the disk and thus the spatial distribution of the lensing event could in principle be used to partially reduce this background. However, since both genuine AML signatures due to PBHs and such background events would peak in the direction of the Galactic center, we expect the spatial distribution of such events to provide only a modest background reduction.

Brown dwarfs and free-floating planets would have masses less than $\sim10^{-2}~M_\odot$. In practice, these objects would be indistinguishable from PBHs of similar mass. As we will see in the next section, our expected exclusion on the PBH parameter space is relevant only for PBH masses $M>10^{-2}~M_\odot$, with lower mass objects giving rise either to SDLEs, or below threshold lensing. Thus, we do not expect brown dwarfs and free-floating planets to be a relevant background.

Lensing signals due to foreground stars form a  reducible background when the foreground star can be resolved by \Gaia{} and thus the source of lensing can be clearly identified. In such a case, either a portion of the trajectory or the entire trajectory could be excluded from an analysis. However, if the foreground star is too dim, for example if it is a black hole or a neutron star or a white dwarf formed from stellar collapse, then lensing due to such objects would form an irreducible background. We would need to model the rate of such lensing events and their observable distribution to predict the expected background rate. The rate of AML signals due to unresolved foreground stars can be predicted in a way similar to that in which we predicted the event rate due to PBHs. One would have to perform a similar calculation as in sec.~\ref{sec:LensingProb}, where instead of using the DM density to characterize the lens distribution we would use a model distribution density for the dark lenses in eq.~\ref{eq:P}. In addition, we would have to modify the monochromatic mass function assumption used for the PBHs, and replace it by a distribution based on a mass model for these faint stellar objects.
A recent study along these lines using a population synthesis code was performed in ref.~\cite{2020ApJ...889...31L}. The authors of this study constructed the distribution of a similar set of observables as those for our IDLEs ($\delta \theta_\textrm{max}$ and $t_e$)\footnote{The authors of ref.~\cite{2020ApJ...889...31L} actually used the Einstein ring crossing time which is $\theta_E/\mu$ instead of the event duration $t_e$ as one of the simplified observables. In order to approximately relate the former to the latter, we can multiply it by $\theta_E/\sigma_a(m_G)$, see eq.~\ref{eq:eventduration}.}, but for lensing due to stellar objects. An examination of their results suggests that the main backgrounds that could give rise to  observables in a similar region of parameter space as the PBH induced observable space (compare our fig.~\ref{fig:ObsDist} with fig.~14 in ref.~\cite{2020ApJ...889...31L}), are those which would arise from neutron stars which could mimic IDLE type events, and black holes from stellar collapse which could mimic very long duration LDLE type events\footnote{The authors of this study found that microlensing due to white dwarfs would give rise comparatively smaller astrometric shift due to their lower masses.}. From a naive comparison by eye, it appears that they expect $\sim$ 1 neutron star event which could mimic an IDLE signal. Although this background seems negligible, a detailed study specifically with \Gaia{} observables is needed to estimate this background more precisely.

The second source of astrophysical background  is due to a violation of our assumption of rectilinear motion. Isolated stars are expected to approximately follow rectilinear motion in the Galactic rest frame over the time period of observation of \Gaia{}. However, gravitational potential gradients can alter this expected behaviour. Such gradients can arise due to localized effects, e.g a binary companion, or localized gravitational perturbations (for e.g. due to a nearby star cluster), or they can arise due to steeper global gravitational gradients, such as near the center of the Galaxy.

Of these possibilities, binary companions would probably be the most significant effect, with nearly a third of main sequence stellar systems in the Galaxy expected to be binary systems~\cite{Lada:2006dc}. Furthermore, when accounting for the bias in selection due to brightness, one would expect a significantly larger fraction of the \Gaia{} sources to be binaries. A study by Penoyre, Belokurov, and Evans~\cite{2022MNRAS.513.2437P, 2022MNRAS.513.5270P} which used a subsample of \Gaia{} data which only utilized nearby stars, found that roughly 10\% of stars in their sample showed significant deviation away from the trajectory expected due to rectilinear motion (more precisely rectilinear + parallax). These deviations were expected to be because of binary companions.

However, this number is not necessarily representative of the expected binary background rate to AML signals. In cases where the binary can be resolved and both companions identified, such systems can easily be identified as backgrounds. In the case of unresolved binaries, or binaries where the companion is too faint to be detected, the relevant astrometric observable would be the motion of the center-of-light (centroid). While the center-of-mass of a binary system is expected to follow a rectilinear trajectory, asymmetric mass-to-light ratios of the binary companions will result in a motion of the center-of-light that will in general show deviations from rectilinear motion. In most cases such deviations will exhibit behavior that is different from that of a genuine lensing signal. One could perform a detailed study of simulated trajectories of the centroid of such binaries and compute their expected event observable distributions. It might then be possible to develop additional discriminatory variables (such as periodic features) that could be used to further separate the binary background from signal. It is also important to know what the true type of lensing signals are, corresponding to figs.~\ref{fig:Smoving}, \ref{fig:LSmoving}, and \ref{fig:Lmoving}, in order to understand the level to which binaries can be confused with genuine AML signals.  For example in the case where the signal involves a static source and a moving lens as in fig.~\ref{fig:Lmoving}, the trajectory for IDLEs would involve some period of the centroid being static, and then moving around a closed loop, and then being static again -- this should be easily distinguishable from most types of binary motion. Alternatively, it might also be possible to leverage the relatively crude photometry of \Gaia{} to check if the candidate AML event also displays a simultaneous PML signal. The development of discriminatory variables, and the degree to which reduction in the background can be achieved requires numerical study which we hope to take up in future work.

Thus, we expect that the true binary background rate can be estimated by a combination of a population model to characterize the number of binary systems which satisfy certain basic criteria to constitute a background (such as asymmetry of the mass-to-light ratios, resolvability, etc.), and further studying the expected reduction in the rate once suitable criteria are imposed to identify binary trajectories from the astrometric motion of the centroid.

In case it is found that the predicted binary background is sufficiently small after such a study, and if the number of detected candidate AML events with \Gaia{} is also sufficiently small, follow up observation of the detected candidates (possibly with terrestrial or other space-based telescopes) could be further used to identify binary companions. If the binary companions are in different spectral classes, photometric follow ups could detect intensity peaks in two widely separated bands~\cite{2011Ap&SS.335..105M, 2014AstBu..69..160S, 2015BaltA..24..137C, 2021AJ....161..276M}, thus identifying it as a binary system. Although this technique would have low efficacy, other additional techniques such as spectroscopic measurements (either to detect different spectral types~\cite{2018MNRAS.473.5043E} or time-domain spectroscopy to measure Doppler shifts~\cite{2002Ap&SS.280..133D, 2004AJ....127.1187T, 2021AJ....162..117B}) or high resolution imaging~\cite{1997ApJ...490..353G, MCALISTER1991123, 2016ApJ...821...52K}, can be used to identify binary systems among candidate lensing events.

We can also estimate the typical time-scale for deviations in trajectories induced by the gravitational potential in a stellar cluster, by computing the ratio of velocity and gravitational acceleration of a star in a typical cluster. Assuming that $v = 200$~km/s as before, and a cluster mass of $10^6~M_\odot$, with a distance of 100~pc between the star and the cluster, we would find a time scale of $\sim 10^8$~years. Therefore, such gravitational gradients would typically only mimic very long duration LDLE type events. Such anomalous accelerations will not exhibit any photometric brightening signals, and this could perhaps be one way to separate them from genuine AML signals.

We can briefly summarize the preceding discussion on astrophysical backgrounds as follows. The potentially largest source of background is due to binary motion. However, a population synthesis modelling, plus a study of the detailed trajectories is needed to estimate the true background rate due to binaries. This background could potentially be further reduced by follow up observations of candidate lensing events. In addition, for very long duration LDLEs we also expect some background from lensing due to black holes from stellar collapse, as well as possible effects of local gravitational gradients on stellar trajectories.

\section{\Gaia{}'s expected exclusion on the PBH parameter space}
\label{sec:Results}
For a given value of $(f,M)$, we have shown that we can calculate the expected number of IDLE and LDLE type events that \Gaia{} will be able to detect. We have also given a procedure to calculate the distributions of some simplified event observables. As discussed in sec.~\ref{sec:background}, there are various types of backgrounds which can mimic genuine PBH induced lensing signals. Given the distribution of expected observables for both signal and background, one could set cuts on the observable parameter space to reduce the amount of background, and maximize the signal-to-background ratio. Then, assuming that \Gaia{} takes $t_\textrm{obs}$ worth of time-series data and sees events consistent with a background-only rate, we could then find the expected exclusion on the regions of PBH parameter space that \Gaia{} can set due to non-observation of any excess event rate over background.

In this work we have only calculated the signal rate and the observable distribution from signal, however we have not computed the full background rates and the observable distributions for all the backgrounds. Thus, we can not give a fully accurate projected exclusion on the PBH parameters that can be set by \Gaia{}.

In order to get some indication of \Gaia{}'s expected sensitivity in the absence of a full background rate calculation, we can make some simple assumptions on the background rate and calculate the projected exclusion. We will assume that we only study AML events with event durations larger than 2 years in order to reject the IDLE statistical background (see sec.~\ref{sec:background_stat}). We will further make the simplified assumption that after setting this cut, the total background from all sources is negligible in the region of observable space where we expect genuine PBH induced AML signatures.

Thus, with the above assumptions, assuming that we see zero candidate lensing events, consistent with background only, we expect to rule out regions of the PBH parameter space that predict $N_l>2.3$ events at the 90\% confidence level, assuming Poissonian statistics.

\begin{figure}
    \centering
    \includegraphics[scale=0.9]{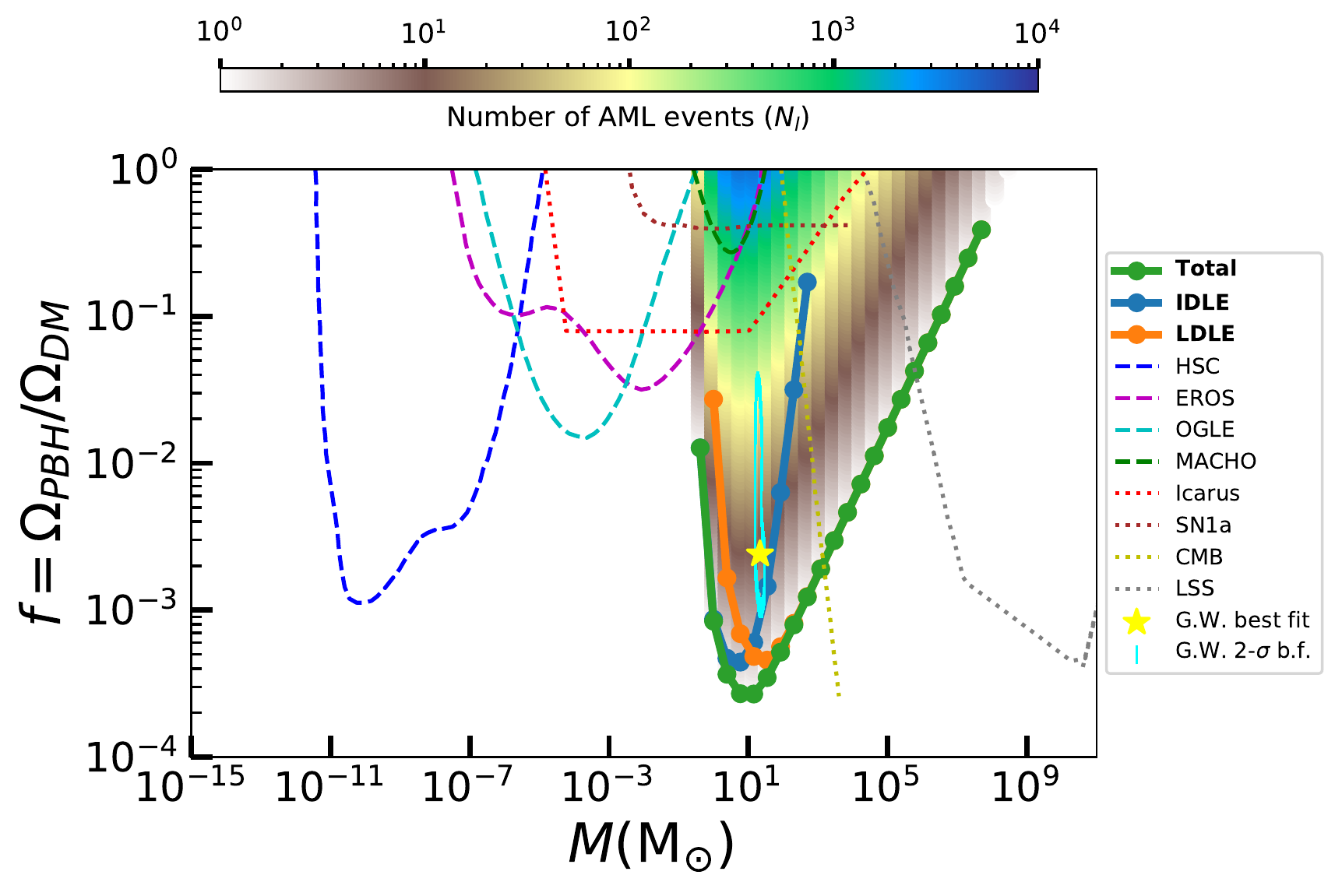}
    \caption{This plot shows excluded regions of the PBH parameter space $(f,M)$. Dashed curves indicate the exclusions that have already been set using PML techniques (HSC~\cite{Niikura2019,Smyth:2019whb}, MACHO~\cite{MACHO:2000nvd}, EROS~\cite{EROS-2:2006ryy}, OGLE~\cite{Niikura:2019kqi}) and dotted curves indicate constraints from a variety of other observations, such as PML signals in giant arcs~\cite{Oguri:2017ock}, lensed SNe1a~\cite{Zumalacarregui:2017qqd}, distortions of the CMB~\cite{Ali-Haimoud:2016mbv}, and effect on large-scale-structure (LSS)~\cite{Carr:2018rid}. The green curve is our result for the expected $90\%$ C.L. exclusion curve that \Gaia{} will set using astrometric microlensing with $t_\textrm{obs} = 5$~years of data. This exclusion curve is computed assuming that \Gaia{} will see no events, consistent with our simplified assumption of a negligible number of expected background events. The total exclusion expected from \Gaia{} can be decomposed into an exclusion based on IDLE type events (blue curve) with event durations between 2 - 5 years, and LDLE type events which have longer event durations. At masses lower than $\sim 0.4~M_{\odot}$, \Gaia{} loses sensitivity to PBHs because such low mass PBHs would yield small deflections and event durations below \Gaia{}'s sensitivity thresholds. At large masses, the number of lenses in the Galaxy decreases and hence the lensing probability also decreases, once again yielding a loss in sensitivity near $\sim10^{8}~M_\odot$. From the curve we see that \Gaia{} is expected to be most sensitive to PBHs with mass $\sim10~M_\odot$. Our projection indicates that  \Gaia{} is sensitive enough to rule out a PBH origin for the LIGO/Virgo gravitational wave (G.W.) signals (indicated by the 2-$\sigma$ best-fit region shown by the cyan contour~\cite{Vaskonen:2019jpv}).}
    \label{fig:exclusion}
\end{figure}

In fig.~\ref{fig:exclusion}, we show the PBH parameter space $(f,M)$ and shaded contours which indicate the average total number of lensing events $N_l$ due to PBHs, for both IDLE and LDLE types, for each point in the parameter space. The boundary of these contours is the green curve, which corresponds to $N_l =2.3$. The region of shaded parameter space interior to the green curve can be excluded at the 90\% confidence level.

We have also shown on the same plot, the curves corresponding to regions of parameter space that give rise to 2.3 IDLE type events (blue curve), or 2.3 LDLE type events (orange curve). The regions above these curves can be interpreted as expected exclusions purely due to null search results for either one of these classes of lensing events.

We can see from the figure that \Gaia{} is sensitive to PBHs with masses between $0.4~M_\odot$ - $5\times10^7~M_\odot$. \Gaia{} is most sensitive to PBH masses near $10~M_\odot$ where a fraction $f$ as low as $3\times10^{-4}$ of the DM relic density can be ruled out. For PBH masses lower than $10^3$~$M_\odot$, most of the detectable lensing events would correspond to IDLEs with average event durations that lie between 2~-~5 years. For sufficiently low PBH masses, two conditions cause the number of detectable lensing events to drop rapidly to zero. Firstly, the Einstein angle $\theta_E$ decreases to such an extent that many of the lensing signals would fall below \Gaia{}'s detection threshold. Secondly, the average event duration $t_e$ would decrease below 2 years, for which the statistical backgrounds to IDLEs would be large. This makes the exclusion bound weaker for low PBH masses. For PBH masses greater than $10^3 M_\odot$, most of the detectable lensing events would correspond to LDLE lensing with average event duration much greater than observation time $t_{\textrm{obs}}< t_e $. For larger PBH masses, the number of lenses decreases with increasing PBH mass, and thus relatively fewer signal events are expected. This makes the bound weaker for larger values of the PBH mass $M$.

We have shown several other exclusion bounds from the literature in fig.~\ref{fig:exclusion}, which can be compared against our projected exclusion bound. Among the existing constraints, lensing constraints are the most robust, since they rely on fewer assumptions. We find that the AML bound on PBH density fraction $f$ that can be set by \Gaia{} for masses $\gtrsim~0.4~M_\odot$ is stronger than that of the current best bounds from PML surveys, possibly by several orders of magnitude. We have also shown in the figure, a best-fit region of the PBH parameter space corresponding to PBH mergers giving rise to all the observed gravitational wave signals seen at LIGO/Virgo~\cite{Vaskonen:2019jpv}. We find that our projected exclusion could rule out this region, thus ruling out PBHs as the origin for the observed gravitational wave events.

A more realistic study of a possible PBH origin for the LIGO/Virgo signals was performed in refs.~\cite{DeLuca:2021wjr,Franciolini:2021tla}. In these works, the authors studied mixed models with some fraction of the LIGO/Virgo signals arising from astrophysical black hole (ABH) mergers, and some from PBHs. Using a Bayesian analysis, they found that only 20\% of the LIGO/Virgo signals are likely to have a PBH origin. This would imply a best fit fraction $f$ of PBH dark matter which is lower than that of~\cite{Vaskonen:2019jpv} by a factor of $\sim$~5~\cite{DeLuca:2021wjr}. However, the exact fraction would depend on the ABH model which is highly uncertain. We note that the expected \Gaia{} exclusion that we have found is strong enough to rule out even this smaller value of $f$, thus potentially ruling out even the best-fit prediction of~\cite{DeLuca:2021wjr}\footnote{Although no other present day constraints exist on such a low fraction of PBHs in this mass range, future gravitational wave searches for PBH mergers at higher redshift $z \gtrsim 30$ with the Cosmic explorer and Einstein telescopes could rule out a fraction $f\gtrsim 10^{-5}$ of PBHs at $10~M_\odot$ in the next decade~\cite{Ng:2022agi}.}.

\section{Discussion}
\label{sec:discussion}
The computation of the expected exclusion curve that \Gaia{} will set based on the AML technique crucially depended on the calculation of the lensing probability $P_\textrm{star}$. There are a number of assumptions that we made while calculating $P_\textrm{star}$, some of which were made for convenience of calculation. We list some ways in which the calculation can be improved.
\begin{itemize}
    \item We made an assumption of a constant relative tangential separation velocity $v = 200$~km/s. This can be further improved by taking into account the stellar velocities and DM velocities. Stellar velocities can potentially be obtained from the \Gaia{} catalog itself, whereas the DM velocities can be obtained perhaps from Eddington inversion techniques of the Galactic rotation curve~(see for e.g. \cite{Mandal:2018efq}). A more accurate prediction of relative velocities is needed because our prediction of the typical event durations (and hence separability from background) has a direct inverse dependence on the separation velocity.
    \item We made the assumption of the rectilinear motion of target stars. However, the parallax motion of these stars (and PBHs) would result in non-rectilinear motion. Moreover, a large fraction of stars in our Galaxy are in binary systems and this would result in wobbles in the star's trajectory due to its bound motion. In order to more carefully take into account these effects in the computation of $P_\textrm{star}$, we would need to perform a numerical simulation of trajectories to study the effects of lensing and how it can be distinguished from other kinds of motion.
    \item We have also assumed uniformly spaced sampling in time by \Gaia{}. However, in a numerical study one could also take into account \Gaia{}'s scanning law which would give us the correct time sampling rate for stars in a given region of sky.
    \item For simplicity, we have taken the astrometric error $\sigma_a(m_G)$ to be the error along the scanning direction of \Gaia{}. In a numerical study, a better estimate of $P_\textrm{star}$ could be obtained using both the across and along astrometric errors depending upon \Gaia{}'s scanning strategy.
    \item The probability $P_\textrm{star}$ depends upon the assumed DM density profile. We computed $P_\textrm{star}$ for the NFW profile, but we could also try other profiles such as Burkert~\cite{Burkert:1995yz},  Einasto~\cite{Graham:2005xx,Navarro:2008kc}, isothermal~\cite{Begeman:1991iy, Bahcall:1980fb} etc. These DM density profiles differ from one another near the Galactic center, and in general are less cuspier than the NFW profile. Since the Galactic center region was where we have found the maximum probability for detecting lensing events, we therefore expect that these alternate profiles could have a sizeable effect on our estimate of the expected number of AML events.
    \item We have ignored blending effects due to foreground stars which could affect the centroiding precision when detecting AML events. A realistic simulation is needed to estimate the size of such effects.
    \item We have assumed that our exclusion is set by using only the AML signature detected by \Gaia{}. However, \Gaia{}'s relatively crude photometry could also be used in conjunction with the AML measurements to measure a PML signal which could potentially improve the sensitivity of the search.
\end{itemize}

In addition to improvements in the signal rate calculation, a detailed background study is needed to obtain an accurate exclusion curve. In particular, numerical studies of the LDLE statistical background are important as this could potentially affect the expected bound at high PBH masses where we expect to have the longest event durations. Also, as outlined in sec.~\ref{sec:background_astro}, we expect that the astrophysical backgrounds that need a more careful study are due to i) wobbles in a stellar trajectory due to an unresolved binary companion or local gravitational potential gradients and  ii) lensing due to dark stellar objects.

\section{Summary: Main takeaway}
\label{sec:Summary}
\Gaia{}'s unprecedented astrometric precision has ushered in a new era in our understanding of the Galaxy. With its milliarcsecond level precision in a single pass, \Gaia{} has the capability to be extremely sensitive to any unusual proper motion of stars. Primordial black holes are expected to form a significant fraction of the dark matter density of the universe in many cosmological models. PBH dark matter in our own Galaxy could lead to astrometric microlensing signals of background stars which could be detected by \Gaia{}.
In this work, we have attempted to make a prediction for the expected sensitivity of \Gaia{} to the PBH parameter space ($f,M$), and to compute the expected AML event observable distributions from PBH induced lensing.

In order to estimate the potential exclusion limit that \Gaia{} can set on the PBH parameter space with 5 years of observational data, we needed to estimate both the rate of genuine PBH induced AML signals, as well as the rate of background events that could mimic such signals.

The present work has primarily focused on a precise prediction of the signal rate and associated observables. In order to compute the signal rate, we used the existing \Gaia{} eDR3 catalog as a model of the stars in the Galaxy which are potential AML targets. We then combined this with a novel probability calculation to estimate the likelihood that a given star in the \Gaia{} catalog would undergo an AML event which is detectable at \Gaia{}. While estimating this probability we argued that there would be three different classes of detectable lensing events, Short Duration Lensing Events (SDLEs), Intermediate Duration Lensing Events (IDLEs), and Long Duration Lensing Events (LDLEs). We also suggested appropriate simplified observables that can be extracted from the apparent trajectory, which characterize the signatures of AML events. For a given set of PBH parameters, our calculations showed a)~the regions of the sky which are likely to yield a large event rate and b)~the distribution of the simplified observables for both IDLE and LDLE types of lensing events.

Although a full background study was beyond the scope of this work, we have also discussed various sources of background, and highlighted which backgrounds we expect to be important and deserving of further investigation. We did perform a slightly more quantitative estimate of the background for IDLE type events due to statistical fluctuations induced by centroiding uncertainties. We found that event durations greater than $\sim 2$~years, would show negligible background of this type. Thus only IDLEs of event durations between 2~-~5 years can be used to search for PBHs, whereas shorter duration IDLEs would not be usefull for PBH searches. We used this to also argue that similarly SDLEs would also be completely swamped by the same statistical background and thus irrelevant for PBH searches. We also highlighted the major sources of expected astrophysical backgrounds and possible techniques that could be used to reduce these backgrounds.

We then computed the expected exclusion curve for \Gaia{} using our predicted signal rate, and the simplified assumption of a negligible background rate. With this assumption, we found that \Gaia{} is sensitive to PBHs with mass between $0.4~M_\odot$ to $5\times 10^7~M_\odot$  with peak sensitivity to PBH masses of $10~M_\odot$, for which we can rule out such PBHs as making up as little as a fraction $f =3 \times 10^{-4}$ of the DM density. The lower end of the mass sensitivity window is due to IDLE type events which yield event durations close to 2 years, near where the statistical background becomes a significant effect. At the higher end of the mass range of interest, \Gaia{} is sensitive to LDLE type events, but loses sensitivity at high masses because the number of lenses decreases, hence decreasing the expected event rate.

As compared to other existing bounds on the PBH parameter space, we have found that \Gaia{} will potentially have the best sensitivity to PBHs with masses near $10~M_\odot$. This region of PBH masses is particularly intriguing since it overlaps with the range of PBH masses that current gravitational wave detectors are sensitive to. An exciting implication of our work is that we expect that \Gaia{} can potentially exclude a PBH origin for the LIGO/Virgo black hole merger events.

Our work is the first attempt to combine a detailed signal rate estimation with a rudimentary background assumption to calculate the expected sensitivity of \Gaia{} to the PBH parameter space. Previous works in the literature have estimated the lensing probabilities, but not the total expected event rate or the exclusion curve. Moreover, these studies have only focused on the case of LDLEs. Our work also shows for the first time that IDLE type events can also play a significant role in excluding or discovering regions of the PBH parameter space with \Gaia{}.

We have also discussed a number of ways in which our calculation of the expected exclusion could be further improved. Most of these improvements require detailed numerical studies which we plan to pursue in the future. Once time-series data of \Gaia{} is publicly available, we expect to be able to analyze this data for potential lensing signals, and detection/non-detection of such signals can then be used to place constraints on the PBH parameter space using our expected event rate calculations.

\acknowledgments

We thank the anonymous referee for many useful comments on the draft. We acknowledge helpful discussions and correspondence with Jeffery J. Andrews, Varun Bhalerao, Surhud More, Jan Rybizki, and Sourav Chatterjee. The work of VR was supported by a DST-SERB Early Career Research Award (ECR/2017/000040).

\appendix

\section{Derivation of some expressions for averaged astrometric lensing observables}
\label{sec:appendixObs}

In this appendix, we derive expressions for the simplified event observables defined in sec.~\ref{sec:obs_dist}, averaged over impact parameters.

For IDLEs, the simplified observable is given by the maximum astrometric deflection $(\delta \theta)_\textrm{max}$. We first find the maximum astrometric shift $\delta u$, in $\theta_E$ units, using eq.~\ref{eq:CentroidPos}. For a rectilinear trajectory, we can take the lens-source separation to be of the form, $u = \sqrt{u_0^2 + \xi^2(t)}$, where $u_0$ is the impact parameter of the trajectory and $\xi(t) =\mu t$ is due to to the rectilinear motion of the source.
The behavior of the astrometric shift $\delta u$, as a function of $u$ was shown in fig.~\ref{fig:delta-u}. For $u_0>\sqrt{2}$, the maximum shift will occur for  when $u=u_0$, i.e. $\xi=0$. However, for $u_0< \sqrt{2}$, the maximum shift will occur when $u=\sqrt{2}$, i.e. when $\xi = \pm\sqrt{2 - u_0^2}$.
Hence, the maximum deflection for a given impact parameter is given by,
\begin{align}
    (\delta \theta)_\textrm{max} & = \begin{cases}
        \frac{u_0\theta_E}{u_0^2 + 2} & u_0> \sqrt{2},\\
        \frac{\theta_E}{2\sqrt{2}} & u_0 \leq \sqrt{2}.
    \end{cases}
\end{align}

We can now average this over all viable impact parameters for which the AML signal would be above threshold, $0<u_0<u_+$ (see eq.~\ref{eq:uplusminus}), to obtain
\begin{align}
    \langle  (\delta \theta)_\textrm{max} \rangle & = \theta_E\frac{1 + \log \left[\frac{1}{8}K(K+\sqrt{K^2 - 8})\right]}{K+\sqrt{K^2 - 8}},
\end{align}
where $K=\frac{\theta_E}{\sigma_a(m_G)}$.

For LDLEs, we take the simplified observable to be the absolute value of the relative deflection difference between the start point and the end point of the trajectory, away from the true rectilinear trajectory. This was computed in ref.~\cite{2000ApJ...534..213D} and was found to be,
\begin{equation}
    \Delta_\textrm{LDLE} = \frac{v t_\textrm{obs} }{D_l}\frac{1}{u^2},
\end{equation}
where $u$ is the angular separation of the source and lens, which is assumed not to change significantly over $t_\textrm{obs}$ for an LDLE. The observable $\Delta_\textrm{LDLE}$ will be above \Gaia{}'s astrometric threshold sensitivity $\sigma_a(m_G)$ for $u<u_T$, where
$u_T = \sqrt{ \frac{ v t_{\text{obs} }  }{D_l \sigma_a(m_G)} } $.

If we further take random positions of the source relative to the lens and assume that $u$ varies in the the range $u_\textrm{min} <u<u_T$, we obtain,
\begin{equation}
    \langle \Delta_\textrm{LDLE} \rangle =\frac{\int_{u_\textrm{min}}^{u_T}  \Delta_\textrm{LDLE} d^2u}{\int_{u_\textrm{min}}^{u_T}   d^2u}= \frac{v t_\textrm{obs} }{D_l} \frac{2\textrm{log}(u_T/u_\textrm{min})}{u^2_T - u^2_\textrm{min}}.
\end{equation}
Here, the averaging is performed over an annular patch of the sky with the integration limits defining the inner and outer radii of the annulus. We choose $u_\textrm{min} = \sqrt{2}$, corresponding to the point at which the instantaneous deflection is maximum, to be the lower limit of integration.

\bibliography{astroMLGAIA}
\end{document}